\documentclass[11pt,a4paper]{article}
\usepackage{jheppub,amsmath,  amssymb,slashed,url,bm,textgreek,upgreek}
\usepackage{graphicx}
\usepackage{epstopdf}
 
\def\tt{{\mathfrak t}}

\def\a{a}

\def\c{{\eurm c}}

\def\be{\begin{equation}}
\def\ee{\end{equation}}

\def\hat{\widehat}

\def\h{\widehat}

\def\dd{{\eurm q}}

\def\Sigma{\Sigma}

\def\O{{\mathcal O}}

\def\A{{\mathcal A}}
\def\d{{\mathrm d}}

\def\b{\overline}
\def\R{{\mathbb R}}
\def\C{{\mathbb C}}
\def\U{{\mathcal U}}

\def\[{\bigl [}

\def\]{\bigr ]}

\def\F{{\mathcal F}}

\def\Z{{\mathbb Z}}

\def\h{\widehat}

\def\K{{\mathcal K}}

\def\G{{\mathcal G}}

\def\M{{\mathcal M}}

\def\W{{\mathcal W}}
\def\P{{\mathcal P}}

\def\H{{\mathcal H}}

\def\bar{\overline}
\def\h{\widehat}

\font\teneurm=eurm10 \font\seveneurm=eurm7  \font\fiveeurm=eurm5
\newfam\eurmfam
\textfont\eurmfam=\teneurm \scriptfont\eurmfam=\seveneurm
\scriptscriptfont\eurmfam=\fiveeurm

\font\teneusm=eusm10 \font\seveneusm=eusm7 \font\fiveeusm=eusm5
\newfam\eusmfam
\textfont\eusmfam=\teneusm \scriptfont\eusmfam=\seveneusm
\scriptscriptfont\eusmfam=\fiveeusm

\font\tencmmib=cmmib10 \skewchar\tencmmib='177
\font\sevencmmib=cmmib7 \skewchar\sevencmmib='177
\font\fivecmmib=cmmib5 \skewchar\fivecmmib='177
\newfam\cmmibfam
\textfont\cmmibfam=\tencmmib \scriptfont\cmmibfam=\sevencmmib
\scriptscriptfont\cmmibfam=\fivecmmib

\def\Tr{{\rm Tr}}
\def\la{\langle}
\def\ra{\rangle}
\def\i{{\mathrm i}}
\def\a{{\sf a}}
\def\b{{\sf b}}
\def\c{{\sf c}}
\def\dd{{\sf d}}
\def\obs{{\rm obs}}
\def\E{{\mathcal E}}
\def\dS{{\rm dS}}
\def\max{{\rm max}}
\def\HH{{\rm HH}}
\def\d{{\mathrm d}}
\def\i{{\mathrm i}}
\def\SO{\rm SO}
\def\HH{{\rm HH}}
\def\SU{{\rm SU}}
\def\x{{\sf x}} 
\def\BH{{\rm BH}}

\title{A Background-Independent Algebra in Quantum Gravity}

 \author{Edward Witten}
\affiliation{School of Natural Sciences, Institute for Advanced Study,\\ 1 Einstein Drive, Princeton, NJ 08540 USA}
\abstract{We propose an algebra of operators along an observer's worldline as a background-independent algebra in quantum gravity.  In that context, it is natural to think
of the Hartle-Hawking no boundary state as a universal state of maximum entropy, and to define entropy in terms of the relative entropy with this state.  In the case that the only spacetimes considered correspond to de Sitter vacua with different    values of the cosmological constant, this definition leads to sensible results.
}

\begin{document}\maketitle

\section{Introduction}\label{intro} 

In ordinary quantum field theory without gravity in a spacetime $M$, we can associate an algebra $\A_\U$ of observables to any open set $\U\subset M$.
However, there are a few problems with this notion in the presence of gravity.

The most obvious problem is that in the context of quantum gravity, since spacetime fluctuates, it is in general difficult to describe the spacetime region that one wants
to talk about.   The options are much more restricted than they are without gravity.

A possibly deeper problem concerns background independence.   In ordinary quantum field theory, the algebra $\A_\U$ that we associate to an open set $\U\subset M$
depends on $M$ and $\U$, of course, but it does not depend on the state of the quantum fields.    What would be the analog of that in gravity?   In gravity, the spacetime
that the observer experiences is part of what the fields determine, so an algebra that does not depend on the state of the quantum fields 
should be defined without reference to any particular spacetime. In other words, it should be background independent.   By contrast, anything we define as the algebra
of the observables in a region $\U\subset M$ will depend on the choice of $M$ and $\U$ and is not background independent.

A third problem concerns the question of why we want to define an algebra in the first place.   What is this algebra supposed to mean?    In ordinary quantum mechanics,
an observer is external to the system and we are quite free to make what assumptions we want about the capability of the observer.   In quantum field theory without gravity,
we can imagine an observer who probes a system at will but only in a specified region $\U\subset M$.   That is the context in which it makes sense to consider the algebra
$\A_\U$: it describes the observations of such an observer.  In gravity, at least in a closed universe or in a typical cosmological model, there is no one who 
can probe the system from outside so an algebra
 only has operational meaning
if it is the algebra of operators accessible to some observer living in the spacetime.

In this article, following many others (for example \cite{Unruh}),  we characterize an observer by a timelike worldline and we assume that what the observer can measure
are the quantum fields along this worldline.   As the simplest possible dynamical principle, we assume that the observer worldline is a geodesic.   The model is meant to be
an idealization of our own situation in the universe.     Our worldline is roughly a geodesic. We have no {\it a priori} 
knowledge of the spacetime we live in,\footnote{For an observer who does have some {\it a priori} knowledge of   the global nature and contents  of the universe,
quite different considerations can apply \cite{BP,BP2,Kirk}.}     but   we have
become aware of a vast universe filled with stars, black holes, galaxies, and all the rest,  primarily by measuring the electromagnetic fields in the
immediate vicinity of our worldline.   And our laboratory experiments can likewise be interpreted as more complex measurements of quantum fields along our worldline.

According to the ``timelike tube theorem'' \cite{Borchers,Araki,Stroh,SW,SWtwo,WittenLecture}, 
in quantum field theory without gravity, 
the algebra of operators along a timelike worldline $\gamma$  is equivalent to the algebra of operators in a 
certain open set, its  timelike envelope\footnote{$\E(\gamma)$ is defined as the set of all points in $M$ that can be reached by deforming $\gamma$ through timelike
curves, keeping its endpoints fixed.   Under favorable circumstances, $\E(\gamma)$ coincides with $J^+(\gamma)\cap J^-(\gamma)$, the intersection of the past and future of $\gamma$,
but in general $J^+(\gamma)\cap J^-(\gamma)$ is larger.   In a general quantum field theory (such as a conformal field theory in two dimensions), 
the timelike tube theorem cannot be strengthened to replace ${\mathcal E}(\gamma)$ 
with $J^+(\gamma)\cap J^-(\gamma)$, but in sufficiently nonlinear theories, this may be possible \cite{WittenLecture}.}   ${\mathcal E}(\gamma)$.  
 So the algebra of operators along a timelike geodesic is a reasonable substitute for the algebra of an open set, and makes more sense when gravity is included. 
 
 Of course, in a full theory of quantum gravity, we expect that an observer cannot be introduced from outside but must be described by the theory.   What it means then
to assume the presence of an observer is that we define an algebra that makes sense in a subspace of states in which an observer is present.   We do not try to define
an algebra that makes sense in all states.  

The background to this article is provided in part by recent work on algebras of observables in quantum gravity in certain situations \cite{LL,LL2,GCP,CLPW,CPW,PW,K,JSS}.
Our starting point will actually be to rethink the construction of \cite{CLPW}, which concerned an observer in de Sitter space, from a different point of view.
In that paper, the motivation for including an observer was that, because of the symmetries of de Sitter space, it was not possible to define a sensible algebra of operators
in the static patch without assuming the presence of an observer. Once an observer is present, operators can be ``gravitationally dressed'' to the worldline of the observer
and an algebra of observables in the static patch can be defined.  In a more general spacetime with less symmetry, operators can be gravitationally dressed to features of the spacetime, so
 this motivation to include an observer
does not apply.   Instead, here we postulate the presence of an observer in order to achieve background independence and for other reasons already described.

The organization of this article is as follows.   In section \ref{basics}, we introduce the idea of a background-independent operator product algebra $\A_\obs$ along the observer
worldline.  This involves reformulating the construction in \cite{CLPW} in a background-independent way.
 We also explain the notion of a state of $\A_\obs$.   In particular, any choice of a spacetime $M$, a geodesic $\gamma\subset M$ that is the observer's worldline,
 and a quantum state of the combined 
system consisting of the quantum fields in $M$ and the observer gives a state of $\A_\obs$.   

In section \ref{static}, we explain the special role of the static patch of de Sitter
space as an example of a spacetime where the observer might be living.   In this case, there is a state $\Psi_\max$ of maximum entropy.   Roughly, it describes empty de Sitter
space in thermal equilibrium with the observer.   The fact that empty de Sitter space has maximum entropy is in accord with previous arguments \cite{Maeda,BoussoOne,BoussoTwo,Banks,BanksFischler,BanksOne,BanksTwo,BFTwo,DST,SusskindA,Susskind}.
 Once one has the state $\Psi_\max$ at hand, one can define a density matrix and entropy for any state of an observer  that can be described as an $\O(1)$ perturbation
of empty de Sitter space.    The entropy of such a state agrees  with the usual generalized entropy
\be\label{zelbo}S_{\rm gen}=\frac{A}{4G}+S_{\rm out},\ee
(for semiclassical states such that the generalized entropy can be defined) 
up to an additive renormalization constant, independent of the state \cite{CLPW}.    

 Suppose, however, that the observer lives in another spacetime, perhaps a spacetime with a different
topology, or  another de Sitter vacuum with a possibly 
different value of the cosmological constant, or simply an $\O(1/G)$ perturbation of the original empty de Sitter spacetime.  
If we are able to make a similar analysis of states of the observer
algebra in that other spacetime, we will arrive at a corresponding definition of entropy, naively with an additive renormalization constant appropriate to this new spacetime.

But we are at risk to have a new renormalization constant for every spacetime (or at least every spacetime that is not continuously connected to one we have already considered).
It would be much more satisfactory to be able to define entropy up to an additive constant independent of the spacetime, so that one could compare entropies of observer states  that are associated with different spacetimes.  It might be impossible to avoid an overall additive renormalization constant independent of the spacetime;
this may   be the price to pay for an approach in which one has algebras and
no quantum mechanical pure states.

  With this in mind, we propose in section \ref{noboundary} that the Hartle-Hawking no boundary state $\Psi_\HH$
can be regarded as a universal maximum entropy state.   We explain in what sense this hypothesis leads to a universal definition of entropy for any state of the observer
algebra, up to a universal additive constant independent of the spacetime, at least for closed universes where the definition of the no boundary state makes sense.  This proposal is speculative, but we show that it leads to a sensible answer in at least one interesting case:
 the case that the spacetimes considered correspond to de Sitter vacua with different values of the cosmological constant.

In a background independent sense, the observer algebra $\A_\obs$ is an operator product algebra, not an algebra of Hilbert space operators.   However, any choice
of a spacetime in which the observer is living gives a Hilbert space representation of $\A_\obs$.   Given such a representation, it is
 possible to complete $\A_\obs$ to a von Neumann algebra $\hat\A_\obs$, 
and one can ask
what sort of von Neumann algebra one gets.   If the spacetime region causally accessible to the observer includes a complete Cauchy hypersurface, then one expects that $\hat\A_\obs$
is of Type I.    In some special cases that are under good control, like the static patch in de Sitter space, one can argue that one gets an algebra of Type II.   It is tempting to conjecture
that $\hat\A_\obs$ is always of Type I or Type II, not Type III, so that the experience of the observer can always be described by a density matrix. 
It is argued heuristically in section \ref{algebras} that this is the case if the no boundary state can indeed be interpreted as a universal state of maximum entropy.

Up to this point in the paper, we consider    an observer who lives inside the spacetime, as opposed to an observer who can probe  spacetime from outside.  
In section \ref{AAdS}, we look from a somewhat similar point of view at asymptotic observables in an asymptotically Anti de Sitter (AAdS) spacetime, which can be probed
from outside.    A large $N$ algebra
of single-trace operators has been studied in several recent papers
\cite{LL,LL2,GCP,CPW}.   However, to define a background independent algebra of single-trace operators, one has to take the large $N$ limit in a somewhat
different way, dividing the single-trace operators by $N$ instead of subtracting their expectation values, in order to get operators that have a  limit for large $N$.   
The result in the large  $N$ limit is a Poisson algebra -- a commutative
algebra,  endowed with a Poisson bracket.   Perturbation theory in $1/N^2$ deforms the Poisson algebra into a noncommutative but associative algebra.
This is the setting of deformation quantization \cite{Sternheimer,WL,Fedosov,Kontsevich,CF,CF2}.   In the present problem, the Poisson algebra can be viewed as an
algebra of functions on any one of the possible classical phase spaces of this problem, which are labeled by the choice of a bulk topology, possibly with additional 
asymptotic boundaries apart from the one on which the algebra is defined.  The noncommutative algebra that arises in the $1/N$ expansion is background independent, but,
like $\A_\obs$, it does not have any preferred Hilbert space representation.   Any choice of a point in any one of the possible classical phase spaces determines such
a representation.  This is analogous to the fact that any spacetime in which the observer might be living determines a Hilbert space representation of $\A_\obs$.

 \section{A Background-Independent Operator Product Algebra}\label{basics}
 
Consider an observer whose worldline is a timelike geodesic $\gamma$ in a spacetime $M$.   First let us discuss the operators along $\gamma$ for the case that $M$ is a fixed
curved spacetime, in the absence of gravity. The worldline is parametrized by the observer's proper time $\tau$.   The observer measures along $\gamma$, for example, a scalar field
$\phi$, or the electromagnetic field $F_{\mu\nu}$, or the Riemann tensor $R_{\mu\nu\alpha\beta}$, as well as their covariant derivatives in directions normal to $\gamma$.  Let us focus
on a particular observable, say $\phi(x(\tau))$, where $x(\tau)$ is the observer's position at proper time $\tau$, and $\phi(x(\tau))$ is the value of $\phi$ at this spacetime point.   We will
abbreviate this as $\phi(\tau)$.   

When we take gravity to be dynamical, we have to take into account that the same observer worldline can be embedded in a given spacetime in different ways, differing by $\tau\to\tau+
{\rm constant}$.   So $\phi(\tau)$ by itself is not a meaningful observable.   We need to introduce the observer's degrees of freedom and define $\tau$ relative to the observer's clock.

In a minimal model, we describe the observer by a rest mass $m$ and a Hamiltonian
\be\label{obsham}H_{\obs}=m +q, \ee
where $q$, which can be interpreted as the Hamiltonian of the observer's clock, is bounded by $q\geq 0$, so that $m$ is the minimum energy of any state of the observer.
To impose the constraint $q\geq 0$, we should only allow operators that commute with the projection operator\footnote{Here
$\Theta$ is the Heaviside theta function, $\Theta(x)=\begin{cases} 1 & x\geq 0 \cr 0 & x<0\end{cases}$.}
$\Pi=\Theta(q) $ onto states with $q\geq 0$.  If $\O$ is any operator, then $\Pi \O \Pi$ commutes with $\Pi$.  So for example, if $p=-\i\frac{\d}{\d q}$ is canonically conjugate
to $q$, then $e^{-\i p}$ is not an allowed operator, but $\Pi e^{-\i p}\Pi$ is allowed.  

To reproduce the Hamiltonian (\ref{obsham}), the observer action should be
\be\label{obsaction}I_\obs=\int_\gamma\d \tau \left(p \frac{\d q}{\d \tau}-\sqrt{-g_{\tau\tau}}(m+q)\right), \ee
where $\tau$ parametrizes the worldline $\gamma$, and $g_{\tau\tau}$ is the restriction of the spacetime metric $g_{\mu\nu}$ to $\gamma$.  With this
action, the equations of motion say that $\gamma$ is a geodesic, and that $q$ is a constant along $\gamma$.  The action (\ref{obsaction}) is invariant under reparametrizations
of $\gamma$.  The reparametrization invariance can be fixed by defining $\tau$ so that $g_{\tau\tau}=-1$. This condition determines $\tau$ up to an additive constant.

Of course, what we have just described is only the simplest model.   As another example, given a scalar field $\phi$,  we could assume the presence of another term in the observer action:
\be\label{nobsaction}\Delta I_\obs =-\int_\gamma \d t\sqrt{-g_{tt}}\lambda\phi(t), \ee
with a coupling constant $\lambda$.   Then $\gamma$ will  no longer be  a geodesic; the gradient of $\phi$ will provide a  force on the observer.  
One could also elaborate the model so that $q$ would not be a conserved quantity.
However, we will consider the simplest possible model, with Hamiltonian (\ref{obsham}) and  action (\ref{obsaction}).  

As already noted, in the presence of gravity, $\phi(\tau)$ is not a meaningful operator because a spacetime diffeomorphism can shift $\tau$ by a constant.  Let $H_{\rm bulk}$ be the generator
of a bulk diffeomorphism that maps $\gamma$ to itself, shifting $\tau$.   There is no canonical choice of $H_{\rm bulk}$, since we have not specified what the diffeomorphism generated
by $H_{\rm bulk}$  should
do away from $\gamma$, but it does not matter what choice we make, since diffeomorphism generators that act trivially along $\gamma$ will anyway be imposed as constraints
in quantizing gravity.   Taking into account the degrees of freedom of the observer, the constraint operator that we should impose is not $H_{\rm bulk}$ but 
\be\label{zelbigo} \h H = H_{\rm bulk}+H_\obs=H_{\rm bulk}+m+q.\ee

We now want to allow only operators that commute with  $\h H$.   Since
\be\label{elbigo}[H_{\rm bulk},\phi(\tau)]=-\i\dot\phi(\tau),\ee
and $m$ is a $c$-number, we need
\be\label{nelbigo} [q,\phi(\tau)]=\i \dot\phi(\tau). \ee
As $q=\i \frac{\d}{\d p}$, we can satisfy this condition by simply setting
\be\label{helbigo}\tau=p,\ee
or more generally 
\be\label{gelbigo}\tau=p+s,\ee
for a constant $s$.

So a typical allowed operator is $\phi(p+s)$, or more precisely
\be\label{juggo} \h\phi_s=\Pi \phi(p+s)\Pi. \ee
In addition to these operators (with $\phi$ possibly replaced by some other field along the observer worldline), 
there is one more obvious operator that commutes with $\h H$, namely $q$ itself.   So we define an algebra $\A_\obs$ that is generated by the $\h\phi_s$ as well as $q$.  

 This construction  hopefully sounds ``background independent,'' since we described it without picking a background.   However, background independence really depends on 
 interpreting the formulas properly.  We will not get background independence if we interpret $\h\phi_s$ and $q$ as Hilbert space operators.   To get a Hilbert space on which
 $\h\phi_s$ and $q$ act, we have to pick a spacetime $M$ and a geodesic $\gamma\subset M$ on which the observer is propagating. 
 Quantization in this spacetime gives a Hilbert space and  we can interpret $\A_\obs$
 as an algebra of operators on this Hilbert space. 
 But we will not have background independence, since   different pairs $M,\gamma$ will, in general, provide inequivalent representations of the
 same underlying operator product algebra.  To get background independence, we have to think of $\A_\obs$ as an operator product algebra, rather than an algebra of Hilbert space
 operators.\footnote{Somewhat similarly, by characterizing a quantum field theory by universal operator product relations, one can define what it means to consider
 the same quantum field theory in different spacetimes \cite{HW,Fred,HW2}.}

In the absence of gravity, we would characterize the objects $\phi(\tau)$ by universal short distance relations.  For example, in a theory that is conformally invariant at short distances,
with $\phi$ having dimension $\Delta$, we would have
\be\label{univop} \phi(\tau)\phi(\tau')= C(\tau-\tau'-\i\epsilon)^{-2\Delta}+\cdots. \ee
This characterization does not require any knowledge about the quantum state. After coupling to gravity and including the observer and the constraint, the operator product expansion (OPE)
in powers of $\tau-\tau'$ becomes an expansion in  $1/q$; see section \ref{further}.   We characterize $\A_\obs$ purely by the universal short distance or $1/q$ expansion of operator products.   With that
understanding, $\A_\obs$ is background-independent.

By a ``state'' of the observer algebra $\A_\obs$, we mean a complex-valued linear function $\a\to \la\a\ra$, $\a\in\A_\obs$, that satisfies two conditions:\vskip.2cm
\noindent
(1) The function $\a\to \la\a\ra$ is positive, in the sense that for all $\a\in\A_\obs$, $\la\a^\dagger\a\ra\geq 0$.\vskip.2cm
\noindent
(2) This function is consistent with all universal OPE relations.   
\vskip.2cm
\noindent
This somewhat abstract notion of a state is analogous to a definition given in \cite{HW2} for quantum field theory in a fixed curved 
spacetime background.\footnote{In a generic open universe, there is no reasonable Hilbert space that contains all physically sensible states of a quantum field.  
To describe all such states,  given the absence of a suitable Hilbert space, the authors of \cite{HW2} 
characterized the quantum field
by universal, state-independent operator product relations, and then they defined a state of the quantum field to be a linear function on this operator product algebra
that is positive and consistent with the OPE relations.}

This definition of a state of the observer is related in the following way to notions that may be more familiar.   Let $M$ be a spacetime and suppose that the observer
worldline is a geodesic $\gamma\subset M$.   If $\H$ is the Hilbert space that describes the fields in $M$ together with the observer, then $\H$ provides a Hilbert space
representation of the algebra $\A_\obs$.
If 
 $\Psi\in\H$ is any state, then the linear function
\be\label{zelo}\a\to \la\Psi|\a|\Psi\ra \ee
is a state of $\A_\obs$, by the abstract definition.   Conditions (1) and (2) are immediate.   We stress that before picking the pair $M,\gamma$, we do not have a Hilbert space
representation of $\A_\obs$, and it is an OPE algebra, not an algebra of Hilbert space operators.
 
 There is a partial converse to this, given by the Gelfand-Naimark-Segal (GNS) construction of a Hilbert space from a state of an algebra.   Suppose that
 $\a\to \la\a\ra$ is a complex-valued linear function that defines a state of the observer algebra.    Formally define a Hilbert space vector $\Psi_1$ that corresponds to this state, and for every $\a\in \A_\obs$,
 define a new vector $\Psi_\a$, in a complex linear fashion, with $\Psi_{\lambda \a+\mu \b}=\lambda \Psi_\a+\mu\Psi_\b$, for $\a,\b\in\A_\obs,$ $\lambda,\mu\in\C$. $\A_\obs$ acts on this 
 set of states
 by $\a\Psi_\b=\Psi_{\a\b}$.      Define inner
 products among these states by $\la\Psi_\a,\Psi_\b\ra=\la \a^\dagger \b\ra$.   By condition (1) in the definition of a state of $\A_\obs$, these inner products are positive semi-definite.
 Taking a completion and dividing by null vectors, one obtains a Hilbert space $\H$ with an action of $\A_\obs$ and a vector $\Psi_1$ such that $\la\a\ra=\la\Psi_{ 1}|\a|\Psi_1\ra$, for all $\a\in
 \A_\obs$.   Thus every state of the algebra $\A_\obs$ in the abstract sense
 is associated to a pure state in some Hilbert space representation of $\A_\obs$.   What is not clear from this reasoning is the extent to which general Hilbert space 
 representations of $\A_\obs$ are related to pairs $M,\gamma$.   
 
 We conclude this section with several technical remarks.
 \vskip.2cm
 \noindent{\bf Remark 1.}  By definition, states $\a\Psi_1$ are dense in the GNS Hilbert space $\H$.   That means that the GNS Hilbert space describes $\O(1)$ perturbations
 of the input state $\Psi_1$, not perturbations of order $1/G$.   So, for example,  empty de Sitter space and de Sitter space perturbed by a classical electromagnetic field
 with energy of order $1/G$ are described by different GNS Hilbert spaces.   Not coincidentally, they are also described by different Hilbert spaces in ordinary perturbation theory; one
 Hilbert space is obtained by perturbing around empty de Sitter space and one is obtained by perturbing around de Sitter space with the electromagnetic wave present.   Of course,
 nonperturbatively  it may be possible to describe empty de Sitter space and de Sitter space with a strong classical field by the same Hilbert space.  In perturbation theory they
 are different.  \vskip.2cm
 \vskip.2cm
 \noindent{\bf Remark 2.}  Roughly speaking, if a clock has Hamiltonian $q$, the time measured by the clock is the conjugate variable $-p=\i \d/\d q$. (At the classical level,
 the equations of motion derived from the action (\ref{obsaction}) are $-\dot p=1$, showing that $-p$ is the time told by the clock, up to an additive constant.)   However, because of
 the constraint $q\geq 0$, it is not possible to define $p$ as a self-adjoint operator that could be measured.    An example of a self-adjoint operator that can serve as a partial substitute
 is $p^2=-\d^2/\d q^2$, defined by Dirichlet boundary conditions at $q=0$.   This operator is self-adjoint, with a complete set of eigenfunctions $\sin(\lambda q)$, $\lambda>0$.   We
 can also define an operator $|p|=(p^2)^{1/2}$, the positive square root of $p^2$.   This operator measures, informally, the absolute value of the time measured by the observer's clock.
 Its expectation value at time $\tau$, assuming a state $\psi_0(q)$ of the observer at time 0, is 
 \be\label{zelf} \bigl \la \psi_0(q)\bigl|\,e^{\i \tau q}|p|e^{-\i \tau q}\,\bigr| \psi_0(q)\bigr\ra = \bigl \la e^{-\i \tau q}\psi_0(q)\bigl|\,|p|\,\bigr|e^{-\i \tau q} \psi_0(q)\bigr\ra.\ee   For large $|\tau|$, this grows as $|\tau|$ towards either the future or the past, so $|p|$ can serve to measure the observer's proper time  in either the far future or the far past.   However, for $|\tau|\lesssim 1$,
 $ \bigl\la \psi_0(q)|e^{\i\tau q}|p|e^{-\i\tau q}|\psi_0(q)\bigr\ra$ depends very much on the assumed initial state $\psi_0(q)$.   It does not seem that any operator accessible to the observer 
 does better than this.
 \vskip.2cm
\noindent{\bf Remark 3.}
To complete the model really requires a refinement that was discussed in section 2.6 of \cite{CLPW}.
By equipping the observer with a Hamiltonian and in effect a clock, we have made it possible to define ``gravitationally dressed'' scalar operators along the observer's worldline.   
However, to enable the observer
to define and measure operators that carry nonzero angular momentum, such as the electromagnetic field or the Riemann tensor, one needs to equip the observer with an orthonormal frame;
in the simplest model (analogous to assuming that the observer worldline is a geodesic), one can assume that this frame is invariant under parallel transport along the observer's worldline.   The phase space of the observer, in $D$ spacetime dimensions, is then  not
$T^*\R_+$ (where $\R_+$ is the half-line $q\geq 0$ and $T^*\R_+$  is its cotangent bundle) but $T^*\R_+\times T^*{\rm Spin}(D-1)$.   Because
the group ${\rm Spin}(D-1)$ is compact, including the second factor does not qualitatively affect our considerations and we will not include it explicitly in this article.   In the real world, 
we effectively have an orthonormal frame at our disposal, and we use it, for example, in mapping  the positions of stars and galaxies in the sky.

\section{The Static Patch}\label{static}

\subsection{The Maximum Entropy State}\label{maxent}

\begin{figure}
 \begin{center}
   \includegraphics[width=3.9in]{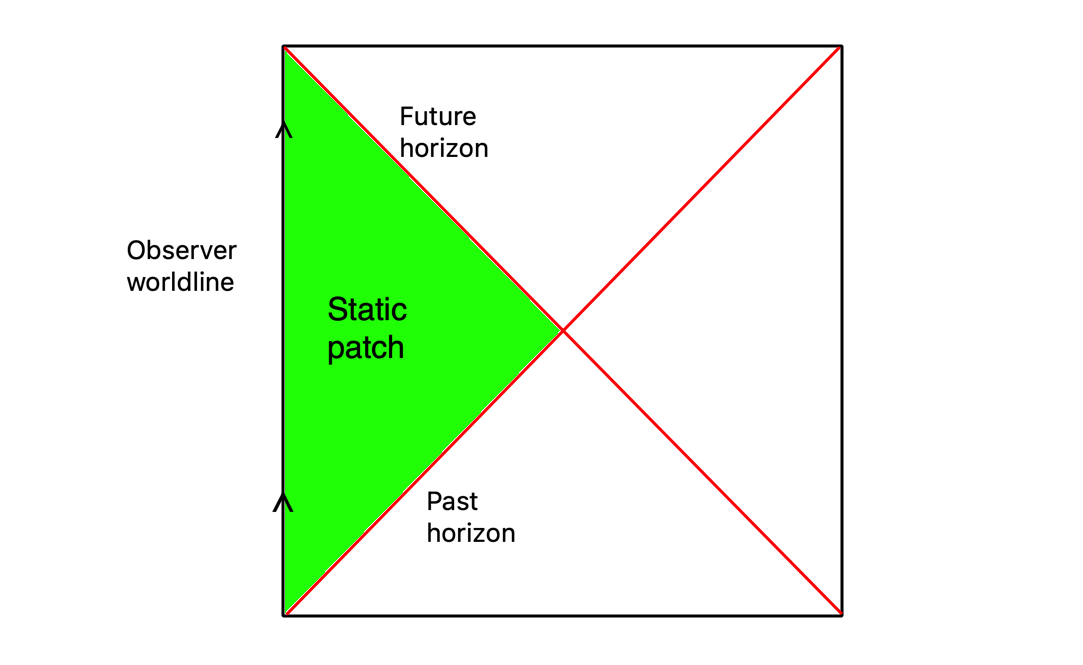}
 \end{center}
\caption{\small \footnotesize{ A Penrose diagram for de Sitter space.  Time flows upward; the far future is at the top of the diagram and the far past is at the bottom.   Coordinates have
been chosen so that the observer's worldline is the left edge of the diagram.   The region causally accessible to the observer is the static patch, which is shaded green.
It is bounded by the past and future horizons of the observer, as shown. }\label{AAA}}
\end{figure}

Any spacetime $M$ in which the observer may be living, together with a choice of geodesic $\gamma$ 
that represents the observer's worldline, leads to a Hilbert space representation of $\A_\obs$.
However, there is a simple special case (previously analyzed in \cite{CLPW} in a way similar to what follows)
that is particularly important.   This is the case that $M$ is an empty de Sitter space, with some positive value of the effective cosmological constant.
De Sitter space in $D$ dimensions has a very large isometry group ${\SO}(1,D)$, under which all geodesics are equivalent, so in this case the choice of $\gamma$ does not matter. 
The region of de Sitter space that is causally accessible to the observer -- the region that the observer can see and also can influence -- is bounded by past and future horizons
\cite{GH}, as indicated in the Penrose diagram of fig. \ref{AAA}.   De Sitter space has a Killing vector field $V$ that  is future directed timelike throughout the causally accessible
region.  It generates a symmetry that maps the geodesic $\gamma$ to itself, shifting it forward in time.   We normalize $V$ so that it looks like $\d/\d\tau$ along $\gamma$,
and we denote the corresponding conserved charge as $H$.   Since  $H$  generates a symmetry that  shifts the observer's
proper time, it can play the role
of the bulk diffeomorphism generator that was called $H_{\rm bulk}$ in the general construction of section \ref{basics}.   
If $H$  is viewed as generating a ``time-translation''
symmetry, then the causally accessible region is time-independent. It has therefore been called a static patch.

In the absence of gravity, quantum fields in de Sitter space have a distinguished de Sitter invariant state $\Psi_{\dS}$ \cite{CT,SS,BD,Mo,Al},  with the property that correlation functions in
this state can be defined by analytic continuation from Euclidean signature.
We normalize this state so that
\be\label{psinor} \la\Psi_\dS|\Psi_\dS\ra=1.\ee
Correlation functions in the state $\Psi_\dS$ are thermal at the de Sitter temperature $T_\dS=1/\beta_\dS$ \cite{GH,FHN}.  It will be helpful to spell out in detail, in the
case of the two-point function of operators $\phi,\phi'$, the meaning of this assertion.   Thermality means that two-point functions
$\la\Psi_\dS|\phi(\tau)\phi'(\tau')|\Psi_\dS\ra$ have two key properties:
\vskip.2cm\noindent
(1) The first property is simply time-translation symmetry,
\be\label{melfo} \la\Psi_\dS|\phi(\tau)\phi'(\tau')|\Psi_\dS\ra=   \la\Psi_\dS|\phi(\tau+c)\phi'(\tau'+c)|\Psi_\dS\ra ,~~~c\in\R.\ee
\vskip.2cm\noindent
(2) The second property is the Kubo-Martin-Schwinger (KMS) condition:
\be\label{kms} \la \Psi_\dS|\phi(\tau-\i\beta_\dS)\phi'(0)|\Psi_\dS\ra =\la\Psi_\dS |\phi'(0) \phi(\tau)|\Psi_\dS\ra.\ee
To be more precise, the KMS condition asserts that the function $\la\Psi_\dS|\phi(\tau)\phi'(0)|\Psi_\dS\ra$, initially defined for real $\tau$, can be analytically continued
to a strip $0\geq {\rm Im}\,\tau\geq -\beta_\dS$, and its values at the lower boundary of the strip satisfy eqn. (\ref{kms}).   The two properties do not hold only for the case that
$\phi,\phi'$ are local operators.   We could, for example, take $\phi(\tau)=\prod_{i=1}^k \phi_i(\tau+s_i)$, with local operators $\phi_i$, and similarly for $\phi'$.  The same statements
 hold without change.

To understand the relation of the KMS condition to thermal equilibrium, consider an ordinary thermal system with Hamiltonian $H$, inverse temperature $\beta$, partition function $Z$,
 and density matrix $\rho=\frac{1}{Z}e^{-\beta H}$.   Time-dependent correlation functions of operators $A,B$ are defined by $\la A(t)B(0)\ra_\beta=\Tr\,\rho e^{\i H t}A e^{-\i H t}B$, 
 $\la B(0) A(t)\ra_\beta=\Tr\,\rho B e^{\i H t}A e^{-\i H t}$, and  from this eqn. (\ref{kms}) immediately follows.  This derivation does not precisely apply to the correlation functions in the
 de Sitter 
 state $\Psi_\dS$, but the KMS condition in that case can be proved using the fact that those correlation functions can be obtained by analytic
 continuation from Euclidean signature.    
 
 Including gravity and the observer, we define a special state\footnote{We have imposed the constraint that operators commute with $\h H$; this constraint is a statement
 about operators that can be defined along the observer's worldline.   One might expect to also impose a constraint that states should be annihilated by $\h H$.
 However, this condition depends on what there is beyond the observer's horizon and places no useful condition on a physical state as a state of the observer algebra 
 $\A_\obs$.  Accordingly we will not discuss such a condition. Such conditions were discussed in \cite{CLPW} and provide no information accessible to the observer.}
 $\Psi_\max$ in which the quantum fields are in the state $\Psi_\dS$, and 
 the observer energy has a thermal distribution at the de Sitter temperature:
 \be\label{psimax}\Psi_\max=\Psi_\dS e^{-\beta_\dS q/2}\sqrt{\beta_\dS}. \ee
 And we replace operators $\phi(\tau)$ by ``gravitationally dressed'' operators $\h\phi_s=\Pi \phi(p+s)\Pi$.   These steps were carried out in \cite{CLPW}, with a somewhat different
 explanation.    Note that by virtue of eqn. (\ref{psinor}), we have
 \be\label{invo} \la\Psi_\max|\Psi_\max\ra=1. \ee
 
 Then a straightforward calculation shows that the two properties (1), (2) that characterize the thermal nature of the state $\Psi_\dS$ are modified as follows:
 \vskip.2cm\noindent
(1$'$) We still have a version of time-translation symmetry, but now it takes the form 
\be\label{melfox} \la\Psi_\max|\h\phi_{s} \h\phi'_{s'} |\Psi_\max\ra=   \la\Psi_\max|\h\phi_{s+c} \h\phi'_{s'+c}|\Psi_\max\ra ,~~~c\in\R.\ee
\vskip.2cm\noindent
(2$'$) The KMS condition simplifies: 
\be\label{tracial} \la \Psi_\max|\h\phi_s \h\phi'_{s'}|\Psi_\max\ra =\la\Psi_\max |\h\phi'_{s'} \h\phi_s|\Psi_\max\ra.\ee
Crucially, there is no shift by $-\i\beta$; the two operators are simply exchanged.   One proof of eqn. (\ref{tracial}) can be found in \cite{WittenLecture}, section 4.   Another proof is
presented shortly in section \ref{traces}.

Condition (2$'$), and its straightforward extension to the additional generator $q$ of $\A_\obs$, tells us that if for $\a\in\A_\obs$, we define 
\be\label{defgo}\Tr\,\a=\la\Psi_\max|\a|\Psi_\max\ra, \ee
then the function $\a\to \Tr\,\a$ does have the algebraic property of a trace:
\be\label{trace} \Tr\,\a\b=\Tr\,\b\a, ~~~\a,\b\in \A_\obs.\ee
This fact   is described by saying that the state $\Psi_\max$ of the algebra $\A_\obs$ is ``tracial.'' 
By virtue of eqn. (\ref{invo}), we have
\be\label{tace}\Tr\,1=1. \ee

Let $\H$ be the GNS Hilbert space of the state $\a\to \Tr\,\a$ of the observer algebra $\A_\obs$.   As explained in section \ref{basics},   this Hilbert space is generated by states
$\a\Psi_\max$, $\a\in\A_\obs$, and (as in Remark 2 at the end of section \ref{basics}) it describes perturbations of the static patch that are of $\O(1)$, not $\O(1/G)$.
$\H$ provides a Hilbert space representation of
$\A_\obs$.  

At this point, we can ask whether a general state $\Psi\in\H$ has a density matrix $\rho$, defined by imitating the standard definition in ordinary quantum mechanics:
\be\label{offput}\la\Psi|\a|\Psi\ra =\Tr\,\a\rho,~~~~~\a\in\A_\obs.\ee
Assuming that the state $\Psi$ is normalized, this condition immediately implies that
\be\label{imly}\Tr\,\rho=1,\ee
as in ordinary quantum mechanics.
By virtue of the definition of the trace, it follows immediately that the state $\Psi_\max$ does have a density matrix, namely $\rho_\max=1$.
We can also easily find the density matrix $\rho_\b$ of a state $\Psi_\b=\b \Psi_\max$, $\b\in\A_\obs$:   from the definitions,  and the tracial nature of $\Psi_\max$, we find that 
$\rho_\b=\b\b^\dagger$
has the desired property $\la\Psi_\b|\a|\Psi_\b\ra=\Tr\,\a\rho_\b$, $\a\in\A_\obs$.   So a dense set of states in $\H$, namely those of the form $\Psi_\b$, 
have a density matrix in $\A_\obs$.

It is not quite true, however, that every state in $\H$ has a density matrix in $\A_\obs$.  $\H$ was defined as a completion of the set of states of the form $\Psi_\b$,
and accordingly, if we want it to be true that every state in $\H$ has a density matrix, we need to replace $\A_\obs$ by a completion $\h\A_\obs$.   This completion,
which is {\it not} background-independent, since it depends on the choice of the Hilbert space representation $\H$, can be defined as the von Neumann algebra
generated by bounded functions\footnote{One passes here to bounded functions of operators in $\A_\obs$, as otherwise it is not clear that the completion is an algebra.}
 of operators in $\A_\obs$.   Once we pass from $\A_\obs$ to its completion, it is true that every state in $\H$ has a density matrix.
To be more precise, every state  $\Psi\in \H$ has a density matrix $\rho$  that is in, or in general affiliated\footnote{An operator $\O$ is affiliated to $\h\A_\obs$ if bounded
functions of $\O$ are in $\h\A_\obs$. 
It is possible for a density matrix $\rho$ to be unbounded (even though it satisfies $\Tr\,\rho=1$).  Since $\h\A_\obs$ is defined as a von Neumann
algebra of bounded operators, an unbounded density matrix is affiliated to $\h\A_\obs$, rather than being contained in $\A_\obs$.} to, $\h\A_\obs$.

Since a dense set of states $\Psi_\b$ have a density matrix $\rho_\b=\b\b^\dagger$ that is manifestly non-negative, it follows that, as in ordinary quantum mechanics,  the density matrix
of any state is nonnegative.   Conversely, if $\rho$ is any non-negative operator in (or  affiliated to) $\h\A_\obs$ satisfying $\Tr\,\rho=1$, 
then it is the density matrix of a state in $\H$, namely
$\Psi_{\rho^{1/2}}=\rho^{1/2}\Psi_\max$.   Indeed, for $\a\in\h\A_\obs$, $\la\Psi_{\rho^{1/2}}|\a|\Psi_{\rho^{1/2}}\ra=\la\Psi_\max|\rho^{1/2}\a\rho^{1/2}|\Psi_\max\ra
=\Tr\,\rho^{1/2}\a\rho^{1/2}=\Tr\,\a\rho$, where in the last step we use the tracial property (\ref{tracial}).   $\Psi_{\rho^{1/2}}$ is exactly analogous to
the  canonical purification of a density matrix $\rho$ in ordinary quantum mechanics. 

Once we know that every state has a density matrix, we can define entropies.   The von Neumann entropy of a state $\Psi$ with density matrix $\rho$ is defined as usual by
\be\label{vonn} S(\rho)=-\Tr\,\rho\log\rho. \ee
In ordinary quantum mechanics, a maximally mixed state is a state whose density matrix is a multiple of the identity, and it has the maximum possible von Neumann entropy.
In the present context, the analog of a maximally mixed state is the state $\Psi_\max$, whose density matrix is $\rho_\max=1$.   
By analogy with what happens in ordinary quantum mechanics,  $\rho_\max$ is a density matrix of maximum entropy.   By the definition (\ref{vonn}), its entropy vanishes:
\be\label{onn}S(\rho_\max)=-\Tr\,1\log 1=0. \ee
On the other hand, every other density matrix has strictly negative entropy.   One way to prove this is as follows.   Let $\rho\not=1$ be some other density matrix,
and for $0\leq t\leq 1$, set $\rho_t=(1-t) +t\rho$.   Then $\rho_t$ is nonnegative and $\Tr\,\rho_t=1$, so $\rho_t$ is a density matrix.   Define $f(t)=S(\rho_t)$.
Then $f(0)=f'(0)=0$, and using the general formula $\log M=\int_0^\infty \d s\left(\frac{1}{s+1}-\frac{1}{s+M}\right)$, one computes
that
\be\label{monn}f''(t)=-\int_0^\infty\d s\,\Tr\,\frac{1}{s+\rho_t}(1-\rho)\frac{1}{s+\rho_t}(1-\rho). \ee
The integrand in eqn. (\ref{monn}) is positive, since it is $\Tr\,L^2$ where $L=(s+\rho_t)^{-1/2}(1-\rho)(s+\rho_t)^{-1/2}$ is self-adjoint. So $f''(t)<0$, $0\leq t\leq 1$.
From $f(0)=f'(0)=0$, $f''(t)<0$, it follows that $f(t)<0$ for $t>0$, and therefore $S(\rho)=f(1)<0$.    

Thus the system consisting of an observer in the static patch has a state of maximum entropy, namely $\Psi_\max=\Psi_\dS e^{-\beta_\dS q/2}\sqrt{\beta_\dS}$, consisting
of empty de Sitter space with a thermal distribution of the observer's energy.  Why did this happen?   The original justification for the claim that empty de Sitter space has
maximum entropy was as follows \cite{BoussoOne}.  Consider a state in which the static patch is not empty, but is filled with particles and fields.    As one evolves to the future, these particles
and fields will all leave the static patch through the future horizon, so the static patch will be empty in the far future.   Since the static patch, from any starting point,
evolves to be empty in the future, the Second Law of Thermodynamics appears to imply that the empty static patch must be a state of maximum entropy.

In the present context, since we have defined the static patch by the presence of an observer,\footnote{As noted in the introduction, a 
possible criticism of approaches that do not explicitly introduce an observer is that, once gravitational fluctuations are considered, 
it is not clear what is meant by the static patch that is under discussion.}       by definition the observer does not leave the static patch 
even in the far future. On the other hand, it is reasonable to expect that in the far future, the static patch will be empty except for the
presence of the observer, and that the observer energy will eventually come into equilibrium with the quantum fields at the de Sitter temperature.\footnote{Actually, if the observer
action is precisely as in eqn. (\ref{obsaction}), this will not happen, since $q$ is a conserved quantity.  A generic small perturbation will ensure that in the far future, the observer reaches
equilibrium with the ambient quantum fields.}   Thus the form of the maximum entropy state $\Psi_\max$ is precisely in accord with what one would expect based on the argument
in \cite{BoussoOne}, once the observer is included.  

A von Neumann algebra (of infinite dimension) that has a trace such that the trace of the identity element is finite -- as in eqn. (\ref{tace}) -- or equivalently,  that has
a state of maximum entropy, here $\Psi_\max$, is said to be of\footnote{This property is not usually taken as the basic definition of a Type II$_1$ algebra.  The usual definition is
explained in \cite{Sorce}, and a simple construction is explained in \cite{Witten}.} Type II$_1$.   So rather as
in \cite{CLPW}, one conclusion is that the algebra $\h\A_\obs$ is of Type II$_1$.  

It is possible to show \cite{CLPW} that for states obtained as $\O(1)$ perturbations of $\Psi_\max$ -- and thus for states in the GNS Hilbert space $\H$ -- 
the entropy defined as in eqn. (\ref{vonn})  agrees, up to an additive constant that is independent of the state, with the usual
generalized entropy
\be\label{sgen}S_{\rm gen}=\frac{A}{4G}+S_{\rm out},\ee
where as usual $A$ is the horizon area and $S_{\rm out}$ is the entropy of particles and fields outside the horizon.  To be more precise, this is true for semiclassical states,
for which the generalized entropy is defined.     The additive constant that is lost in this algebraic definition of entropy is large -- it is the entropy of the maximum entropy
state.   With the definitions that we have given, the maximum entropy state has entropy zero, and all entropies are measured relative to that.
By contrast, in the standard approach \cite{GH}, the entropy of the maximum entropy state is large, approximately $A_\dS/4G$, where $A_\dS$ is the horizon entropy for empty de Sitter space.

Physically, the meaning of the constant discrepancy between the two notions of entropy is that the entropy defined in terms of a state of $\h\A_\obs$ is a sort of renormalized
entropy, from which a renormalization constant has been subtracted.    The maximum entropy state of a Type II$_1$ algebra can be described in terms of an infinite number
of qubits in a maximally entangled state, so its entropy is naturally infinite (for example, see \cite{Witten} for an explanation of this).   
This infinity needs to be renormalized away, but the algebraic approach  via $\h\A_\obs$ does not have enough information to know what 
value to assign to the entropy of the maximum entropy state, and it is usually just set to zero, as we have done in the preceding discussion.   
In section \ref{noboundary}, however, we will try to do  better, at least in comparing different spacetimes.

\subsection{A Proof of The Tracial Property}\label{traces}

Our goal in this section is to prove the tracial property of the maximum entropy state $\Psi_\max$.   Let us first formulate exactly what we wish to prove.
First, let $\a,\b$ be any observables along the observer's worldline in the absence of gravity and without imposing the constraint.
For example, as in section \ref{maxent}, we could have
\be\label{tono}\a=\phi(s),~~~\b=\phi'(s'),\ee
for some scalar fields $\phi,\phi'$ and times $s,s'$.   
But $\a$ and $\b$ could be more complicated; for example, $\a$ could be a product of scalar fields at different times: $\a=\prod_{i=1}^n \phi_i(s_i)$.

After including the observer degrees of freedom and imposing the constraint, we replace $\a$ and $\b$ with gravitationally dressed operators
\be\label{ono}\h\a=\Pi \a(p)\Pi,~~~\h\b=\Pi \b(p)\Pi, \ee
where
\be\label{zono}\a(p)=e^{\i p H}\a e^{-\i p H},~~~\b(p)=e^{\i p H}\b e^{-\i p H}.\ee
Here $H$ is the time translation generator of the static patch.

We now wish to prove that
\be\label{toprove} \la\Psi_\max|\h\a\h\b|\Psi_\max\ra = \la\Psi_\max|\h\b\h\a|\Psi_\max\ra.\ee
In fact, since the algebra $\A_\obs$ has one more generator $q$, we will want to prove a slightly more general statement, as explained later.

Because $H\Psi_\max=0$, $\Pi\Psi_\max=\Psi_\max$, and $H$ commutes with $\Pi$, eqn. (\ref{toprove}) simplifies to
\be\label{oprove}\la\Psi_\max|\a(p)\Pi \b(p)|\Psi_\max\ra =\la\Psi_\max|\b(p)\Pi \a(p)|\Psi_\max\ra. \ee
The structure of eqn. (\ref{oprove}) suggests that it is convenient to describe the observer Hilbert space $\H_\obs$ as a space of functions of $p$,
with $q=\i \frac{\d}{\d p}$.   If we do this, the constraint $q\geq 0$ means that $\H_\obs$ should be defined to consist of square-integrable functions $f(p)$
that are holomorphic and decaying in the lower half $p$-plane.   For example, a state with $q=q_0$ is $\exp(-\i q_0 p)$, and this decays in the lower half-plane
if and only if $q_0>0$. 
The inner product on $\H_\obs$ is the standard
\be\label{innpro}\la f(p)|g(p)\ra =\int_{-\infty}^\infty \d p \,\bar f(p) g(p). \ee

  In this representation, the projection operator $\Pi=\Theta(q)$ from $L^2$ functions on $-\infty<q<\infty $ to $L^2$ functions supported on $q\geq 0$ is an integral operator with the kernel
\be\label{intker}K(p,p')=\frac{1}{2\pi \i}\frac{1}{p-p'-\i\epsilon}. \ee
In fact, by closing the integration contour in the upper or lower half-plane, one can show that
\be\label{nitker}\int_{-\infty}^\infty \d p' \,K(p,p')e^{-\i q p'}=\begin{cases}  e^{-\i q p'} & {\rm if}~q>0\cr 0 & {\rm if}~q<0,\end{cases}\ee
implying that $K(p,p')$ is the integral kernel of the orthogonal projection operator  on $q\geq 0$.

In this representation, the state of the observer that we have formerly written as $\sqrt{\beta_\dS} e^{-\beta_\dS q/2}\Theta(q)$ becomes 
$\sqrt{\frac{\beta_\dS}{2\pi}}\frac{1}{p-\i \beta_\dS/2}.$  Therefore
\be\label{logo}\Psi_\max =\Psi_\dS \sqrt{\frac{\beta_\dS}{2\pi}}\frac{1}{p-\i \beta_\dS/2}.\ee

 Using this expression for  $\Psi_\max$ and expressing $\Pi$ in terms of the kernel $K$, we get
 \be \label{delfic} \hskip-7.4cm \la\Psi_\max|\h\a \h\b |\Psi_\max\ra = \la\Psi_\max|\a(p) \Theta(q) \b(p)|\Psi_\max\ra \ee
\be\notag \hskip2.72cm =\frac{\beta_\dS}{2\pi}\int_{-\infty}^\infty \frac{\d p \,\d p'}{2\pi \i}\frac{1}{
(p+\i \beta_\dS/2)(p-p'-\i\epsilon)(p'-\i\beta_\dS/2)} \la\Psi_\dS|\a(p)\b(p')|\Psi_\dS\ra.\ee
Using time-translation symmetry (\ref{melfo}) and defining $v=p-p'$, this becomes
\be\label{beco}\frac{\beta_\dS}{2\pi}\int_{-\infty}^\infty \frac{\d p \,\d v}{2\pi\i} \frac{1}{(p+\i\beta_\dS/2)(v-\i\epsilon)(p-v-\i\beta_\dS/2)} \la\Psi_\dS|\a(v)\b(0)|\Psi_\dS\ra. \ee
Integrating over $p$, we learn  that
\be\label{intp}\la\Psi_\max|\h\a\h\b|\Psi_\max\ra=  \frac{\beta_\dS}{2\pi}\int_{-\infty}^\infty \d v\frac{1}{(v-\i\epsilon) (v+\i\beta_\dS)} \la\Psi_\dS|\a(v)\b(0)|\Psi_\dS\ra.\ee
Finally we make use of holomorphy and the  KMS property.   The integrand in eqn. (\ref{intp}) is holomorphic in a strip $0\geq {\rm Im}\, v>-\beta_\dS$.  So we can shift the
integration contour by $v\to v-\i\beta_\dS+\i\epsilon$, getting
\be\label{intop}\la\Psi_\max|\h\a\h\b|\Psi_\max\ra=  \frac{\beta_\dS}{2\pi}\int_{-\infty}^\infty \d v\frac{1}{(v-\i\beta_\dS) (v+\i\epsilon)} \la\Psi_\dS|\a(v-\i\beta_\dS)\b(0)|\Psi_\dS\ra.\ee
Using the KMS property (\ref{kms}) and setting $v=-w$, we get 
   \be\notag\la\Psi_\max|\h\a\h\b|\Psi_\max\ra= \frac{\beta_\dS}{2\pi} \int_{-\infty}^\infty \d w \frac{1}{(w-\i\epsilon)(w+\i\beta_\dS)}\la\Psi_\dS|\b(0)\a(-w)|\Psi_\dS\ra
   \ee    
   \be\label{zofo} \hskip3.5cm= \frac{\beta_\dS}{2\pi} \int_{-\infty}^\infty \d w \frac{1}{(w-\i\epsilon)(w+\i\beta_\dS)}\la\Psi_\dS|\b(w)\a(0)|\Psi_\dS\ra,\ee
   where time-translation symmetry was used again.
Comparing to eqn. (\ref{intp}), this implies
  the claimed result
  \be\label{moko}\la\Psi_\max|\h\a\h\b|\Psi_\max\ra=\la\Psi_\max|\h\b\h\a|\Psi_\max\ra.\ee

To complete the picture, we have to take into account that the algebra $\A_\obs$ has one more generator, namely $q$.  A sufficiently rich set of functions of $q$
are the exponentials $e^{\i s q}$ for real $s$.   To complete  the analysis, it suffices
to check the tracial property for operators\footnote{Note that $\h\a_{[s]}$ does not coincide with $\h\a_s=\Pi\a(p+s)\Pi$ as defined in eqn. 
(\ref{juggo}).}
 $\h\a_{[s]} =\h\a e^{\i s q}$, $\h\b_{[s']} =\h\b e^{\i s'q}$, with $\h\a$, $\h\b$ as before and $s,s'\in\R$.  Using the fact that
$e^{\i s q}=e^{-s\frac{\d}{\d p}}$ acts on $p$ by $p\to p-s$, one can repeat the previous steps.
   For example, the generalization of eqn. (\ref{intp}) turns out to be
   \be\label{bintp}\la\Psi_\max|\h\a_{[s]}\h\b_{[s']}|\Psi_\max\ra=  \frac{\beta_\dS}{2\pi}\int_{-\infty}^\infty \d v\frac{1}{(v-s-\i\epsilon) (v+s'+\i\beta_\dS)} \la\Psi_\dS|\a(v)\b(0)|\Psi_\dS\ra.\ee
Again using holomorphy and shifting the integration contour by    $v=v-\i\beta_\dS+\i\epsilon$ and then repeating the previous steps,
we arrive at   
    \be\label{molko}\la\Psi_\max|\h\a_{[s]}\h\b_{[s']}|\Psi_\max\ra=\la\Psi_\max|\h\b_{[s']}\h\a_{[s]}|\Psi_\max\ra.\ee
    This confirms that the thermal property of $\Psi_\dS$ leads, after coupling to gravity and including the observer, to the tracial property of $\Psi_\max$.
    
   The reader may wonder whether in order to complete the proof of the tracial property, we need to prove that $\Tr\, \h\a_{[s]}\h\b_{[s']}\h\c_{[s'']}=\Tr\,\h\b_{[s']}\h\c_{[s'']}\h\a_{[s]}$,
   and similarly with more than three operators.   The answer is that this is not necessary, because in the preceding proof, we did not assume $\a$ or $\b$ to be local operators, and
   any product of the form $\h\b_{[s']}\h\c_{[s'']}$ can actually be expressed as a linear combination of operators $\h\dd_{[s''']}$, for some $\dd$.  To do this, we write
   \be\label{polz}\h\b_{[s']}\h\c_{[s'']}=\Pi \b(p)e^{\i s' q}\Theta(q) \c(p) e^{\i s'' q}\Pi.\ee   Using $\Theta(q)=\int_{-\infty}^\infty\frac{\d\lambda}{2\pi\i}\frac{1}{\lambda-\i\epsilon} e^{\i \lambda q}$,  
   we get 
\be\label{olz} \h\b_{[s']}\h\c_{[s'']}=\int_{-\infty}^\infty\frac{\d\lambda}{2\pi\i}\frac{1}{\lambda-\i\epsilon} \Pi \b(p) e^{\i (s'+\lambda)q } \c(p) e^{\i s''q}\Pi.\ee
Now we write $e^{\i (s'+\lambda)q } \c(p) e^{\i s''q}= \c_{\lambda+s'}(p)e^{\i(s'+s''+\lambda)q}$, leading to
\be\label{nolz}\h\b_{[s']}\h\c_{[s'']}=\int_{-\infty}^\infty\frac{\d\lambda}{2\pi\i} \frac{1}{\lambda-\i\epsilon} \h\dd_{[s'+s''+\lambda]},~~~~ \dd=\b \c_{\lambda+s'}.\ee
This is of the claimed form.
It follows, for example, that states $\h \b_{[s]}\Psi_\max$ are dense in the GNS Hilbert space.  There is no need to add states $\h\b^1_{[s_1]}\h\b^2_{[s_2]}\cdots\h\b^k_{[s_k]}\Psi_\max$
to get a dense set of states.

\subsection{Some Further Properties}\label{further}

In quantum field theory without gravity, what we informally call  a ``local operator'' $\phi(x)$ is not really a  Hilbert space operator, since in acting on a normalizable state it
always produces an unnormalizable state, mapping us out of Hilbert space.   To get a Hilbert space operator, we have to smear $\phi(x)$ in spacetime.   In fact, smearing
along a timelike curve, such as the worldline of an observer, is enough to produce a  Hilbert space operator, 
albeit one that is unbounded
and therefore only densely defined.
This was shown originally (for the case of a geodesic in Minkowski space) in  \cite{BorchersTwo} and has been reviewed recently \cite{WittenLecture}.

After coupling to gravity and introducing the observer, we replace, for example, a local operator $\phi(\tau)$ with a gravitationally dressed version $\h\phi_s$.
One may wonder if $\h\phi_s$, like the underlying $\phi(\tau)$,  requires some smearing to turn it into a true Hilbert space operator.   The answer to this question
is that no smearing is needed; $\h\phi_s$ is already an (unbounded)  Hilbert space operator.  Roughly speaking, gravitational dressing has provided the necessary smearing.

This actually follows from some of the facts that were used in proving the tracial property.
   The two-point function $\la\Psi_\dS|\a(v)\b(0)|\Psi_\dS\ra$ that appears in eqn. (\ref{bintp})  is in general singular on the real
$v$ axis.   But  this two-point function is the boundary value of a function holomorphic in a strip $0>{\rm Im}\, v>-\beta_\dS$.
The function $1/{(v-s-\i\epsilon) (v+s'+\i\beta_\dS)} $ that multiplies this correlation function in eqn. (\ref{bintp}) is holomorphic in the same strip.   Hence we can deform
the integration contour into the middle of the strip, say at ${\rm Im}\,v=-\beta_\dS/2$.   This makes it obvious that the integral that computes
$\la\Psi_\max|\h\a_s\h\b_{s'}|\Psi_\max\ra$ is always convergent, regardless of what we choose for $\a,\b,s$, and $s'$.   This is true even if we arrange so that $\h\a_{s}$
is the hermitian adjoint of $\h\b_{s'}$.   So $\h\b_{s'}\Psi_\max$ is normalizable; it is a Hilbert space state.  

There is a simple explanation of why this has happened, and this will hopefully make it obvious that $n$-point functions of these operators are similarly finite without any need
for smearing.    
Let us consider a two-point function in the absence of gravity in the underlying state $\Psi_\dS$:
\be\label{toffo} G(\tau)=\la\Psi_\dS|\phi(\tau)\phi'(0)|\Psi_\dS\ra. \ee
The function $G(\tau)$  is singular at $\tau=0$.   The singularity comes from a sum over excitations with high energy, that is, with a large eigenvalue of the de Sitter
generator $H$, created by $\phi'(0)$ and then annihilated by $\phi(\tau)$.   However, when we include the observer and impose the constraint, $\phi(\tau)$ and $\phi'(0)$ 
are replaced by operators $\h\phi_s$ and $\h\phi'_{s'}$ that commute with $\h H=H+m+q$, and instead of $G(\tau)$ we consider a dressed correlation function
\be\label{boffo}\h G=\la\Psi_\max|\h\phi_s \h\phi'_{s'}|\Psi_\max\ra. \ee
Since $\h\phi'_{s'}$ commutes with $H+m+q$, in order for it to create an excitation of large $H$, it will have to reduce the value of $q$ by the same amount.   But in the state
$\Psi_{\max}$, it is exponentially unlikely to observe a value of $q$ much greater than $1/\beta_\dS$, and $q$ is strictly not allowed to be negative.   So it is exponentially unlikely 
for $\h\phi'_{s'}$ to  reduce $q$ by much more than $1/\beta_\dS$, and therefore it is exponentially unlikely for $\h\phi'_{s'}$
 to create a state with $H\gg 1/\beta_\dS$.   Hence the sum over high energy states is cut off, and the function $\h G(\tau)$ is finite for any choices of the operators.

This explanation makes it clear that the energy cutoff depends on the choice of the specific state $\Psi_\max$.   Let us consider a more general state $\Psi_f=\Psi_\dS \otimes f(q)$,
replacing the specific function $e^{-\beta_\dS q/2}\sqrt{\beta_\dS}$ that is used in the definition of $\Psi_\max$ with a more general function $f(q)$.   If $f(q)$ is supported at
$q\sim q_0$, then  in eqn. (\ref{boffo}),  the sum over  intermediate states  will be cut off at  $H\sim q_0$.  This corresponds to a short distance cutoff at $\tau\sim 1/q_0$.  So 
for example if in the absence of gravity $\phi$ and $\phi'$ are scalar fields with the property that the most singular term in the operator product expansion is
\be\label{zondo} \phi(\tau)\phi'(\tau')\sim \frac{C}{(\tau-\tau'-\i\epsilon)^{2\Delta} }+\cdots,\ee
then we expect
\be\label{wondo} \la\Psi_f |\h\phi_s \h\phi'_{s'} |\Psi_f\ra \sim q_0^{2\Delta}+\cdots. \ee
It is not difficult to verify this by generalizing slightly the computations in section \ref{traces}.  Since the two-point functions can be arbitrarily large, depending on $f$, the 
operators $\h\phi_s$ are unbounded.

More generally, the usual short distance expansion in decreasing powers of $1/(\tau-\tau'-\i\epsilon)$ becomes, after including the observer and 
coupling to gravity,  a high energy expansion in decreasing powers of $q$.

In a state of the form $\Psi_f=\Psi_\dS\otimes f(q)$ where $f(q)$ is supported at $q\cong q_0$, one will have 
\be\label{zolb} \la\Psi_f|\h\phi_s\h\phi'_{s'}|\Psi_f\ra \cong \la \Psi_\dS|\phi(s)\phi'(s')|\Psi_\dS\ra\ee
for $|s-s'|\gg 1/q_0$, since under that restriction the projection operators $\Pi$ in the definition of $\h\phi_s$ and $\h\phi'_{s'}$ will not play an important role.
In other words, two-point functions will satisfy an approximate equality (\ref{zolb})  if the proper time separation between the two operators is much greater than $1/q_0$.  To the
extent that the relation (\ref{zolb}) holds, the observer is able to see ordinary physics in the underlying de Sitter space.   
In the case of the state $\Psi_\max$, one has $q_0\sim 1/\beta_\dS$, which is the time scale of the exponential expansion of de Sitter space.   Thus in the state $\Psi_\max$,
an approximate equality (\ref{zolb}) does not hold at sub-cosmological time scales.   The  relation (\ref{zolb}) does approximately
hold in the state $\Psi_\max$ on super-cosmological time scales, but does not contain much information, since on such time scales, the two-point functions reduce to products of one-point functions.

One can slightly modify the model under discussion by assuming an upper bound as well as a lower bound on the value of $q$.  This actually makes the model more realistic:
the upper bound on $q$ is the total energy available to the observer in performing any experiment.      Everything that has been said up to this point remains
valid if $q$ is bounded above as well as below.     As usual, an upper bound on the energy available for an experiment places a lower
bound on the time scales that the experiment can resolve.     In a model with an upper bound $q\leq q_*$, the observer can resolve ordinary nongravitational physics down to
a time scale of order $1/q_*$.  

Going back to the question of defining the $\h\a_s$ as Hilbert space operators, the fact that $n$-point functions of such operators are finite implies,
for example, that states of the general form   $\h\a_s\h\b^1_{s_1}\h\b^2_{s_2}\cdots \h b^k_{s_k}\Psi_\max$ are  normalizable (here $\a$ and $\b^1,\cdots,\b^k$ are operators
on the observer worldline in the absence of gravity, and $s,s_1,\cdots,s_k$ are real parameters).   
This implies that the  $\h \a_s$ can  be defined as operators on the GNS Hilbert space that have
a common dense domain  consisting of states of the form $\h\b^1_{s_1}\h\b^2_{s_2}\cdots \h b^k_{s_k}\Psi_\max$.  In view of what is explained at the end of section \ref{traces},
 states $\h\b_s\Psi_\max$ are actually sufficient to comprise a (slightly smaller) dense domain.

\section{Entropy And The No Boundary State}\label{noboundary}

\subsection{$\Psi_\dS$ and $\Psi_\HH$}\label{bothstates}

In section \ref{static}, we considered an observer in the static patch in de Sitter space.    For any state that can be described as an $\O(1)$ perturbation of the empty
static patch, we gave a definition of entropy.   This definition suffers from the need for an arbitrary renormalization constant, but up to an additive constant that is independent
of the state, it seems to be a satisfactory notion in the sense that it agrees with previously known definitions of gravitational entropy when they are available \cite{CLPW}.

Now suppose we consider the observer living in a different spacetime.   A different spacetime might be topologically different, or it might be derived from a different de Sitter vacuum
of the same underlying theory, or we might simply consider an $\O(1/G)$ (rather than $\O(1)$) perturbation of the original de Sitter space.   If we are successful in adapting the analysis
of section \ref{static} to a different spacetime, we will again get a definition of entropy for any state obtained as an $\O(1)$ perturbation of this spacetime, again up to an additive
constant.

It is not very satisfactory to have a new renormalization constant for every new spacetime that we consider, especially because one suspects that (at least among closed universes,
or among spacetimes with a common asymptotic behavior at spatial infinity) at a nonperturbative level, the different spacetimes are all continuously connected.  It would be much
nicer to find a definition of entropy subject only to a single overall additive renormalization constant, independent of the spacetime.   Then we could compare different spacetimes.
We will propose such a definition here, at the cost of going somewhat out on a limb.  One overall renormalization constant may be the price of a semiclassical approach based on algebras
rather than quantum mechanical microstates.

First we will take advantage of the existence of a maximum entropy state to
reinterpret entropy in terms of relative entropy.   We recall that in ordinary quantum mechanics, the relative entropy between two density matrices $\rho$ and $\sigma$
is defined as
\be\label{turnof}S(\rho|\sigma)=\Tr\,\rho(\log\rho-\log\sigma). \ee
Clearly, this definition makes sense for the algebra $\h\A_\obs$ of the static patch, since this algebra has a trace and a notion of density matrices.
The static patch algebra has a state $\Psi_\max$ of maximum entropy, with density matrix $\rho_\max=1$.   From the definition, we see immediately
that, since $\log\,\rho_\max=0$,  the entropy $S(\rho)$ of any density matrix can be expressed in terms of its relative entropy\footnote{A previous paper in which it has
been useful to understand entropy as relative entropy with a special state was \cite{Wall}, in which the generalized entropy outside a black hole horizon was interpreted
as relative entropy with the Hartle-Hawking state of the black hole.   This was a step in proving the Generalized Second Law.}   with the maximum entropy state:
\be\label{urnof} S(\rho)=-S(\rho|\rho_\max). \ee

In ordinary quantum mechanics with a Hilbert space of dimension $N<\infty$, something similar is true, but with an additive constant independent of the state.
In that case, the density matrix of maximum entropy is $\rho_\max=1/N$ and instead of eqn. (\ref{urnof}), we have
\be\label{zurnof}S(\rho)=-S(\rho|\rho_\max)+\log N. \ee
A Type II$_1$ algebra can be viewed as a large $N$ limit of ordinary quantum mechanics, with the states considered being almost maximally mixed,
and with entropy defined with an additive renormalization that removes the additive constant $\log N$ and sets the entropy of the maximum entropy state to vanish.  See \cite{Witten}
for more detail on this.

The de Sitter invariant state $\Psi_\dS$ with which we began our discussion of the static patch can be obtained by analytic continuation from Euclidean signature.
Let us spell out what that means.   The Euclidean analog of de Sitter $D$-space is a $D$-sphere $S^D$.   The ``equator'' of the sphere is a $(D-1)$-sphere $W$,
which can be viewed as the boundary of the southern (or northern) hemisphere $H$.  The state $\Psi_\dS$ can be understood as a state of quantum fields on $W$
that is obtained by a path integral on $H$, keeping fixed the boundary values on $W=\partial H$.    To be precise, suppose for example that we are studying a scalar field $\phi$.
We will write $\phi_W$ for a classical $\phi$ field defined on $W$ and $\phi_H$ for such a field defined on $H$.  A state of the quantum fields on $W$ in this model is
a function $\Psi(\phi_W)$.   The particular state $\Psi_\dS(\phi_W)$ can be found by a path integral over $\phi_H$ subject to the condition that $\phi_H|_W=\phi_W$ (here $\phi_H|_W$ is
the restriction of $\phi_H$ to $W$):
\be\label{zilno}\Psi_\dS(\phi_W)=\int_{\phi_H|_W=\phi_W}D \phi_H \exp(-I(\phi_H)). \ee

The Hartle-Hawking no boundary state \cite{HH}, which we will call $\Psi_\HH$, is based on a similar idea in the context of gravity.   
To adapt the definition of $\Psi_\dS$ to gravity, one of the fields on which the wavefunction depends should be 
a metric $g_W$ on $W$. Also in a theory of gravity, one has to sum over all possible choices of manifolds $H$ with $W=\partial H$, rather than just choosing one, as in the
definition of $\Psi_\dS$.   This leads to the definition of $\Psi_\HH(g_W)$ as a path integral over all manifolds $H$ of boundary $W$; one sums over the choice of $H$, and
 for each $H$, one integrates over the metric
$g_H$ on $H$, with the restriction $g_H|_W=g_W$.   If other fields are present as well, they are included in an obvious way: one formally defines the no boundary state
$\Psi_\HH(g_W,\phi_W,\cdots)$ by summing and integrating over all bulk data that restrict to the given boundary data on $W$.  The state is called a no boundary state because
spacetime is taken to have no boundaries except a specified boundary on which the quantum state is defined.  The rest of this section will be a brief review of aspects of the no boundary
state and an explanation of its extension to include an observer.

For a variety of reasons, including the fact that the Einstein action in Euclidean signature is unbounded below, there are many unanswered questions about the
 no boundary state.   Everything
about it can be questioned.     However, 
assuming the cosmological constant is positive so that a $D$-sphere of appropriate radius
is a classical solution,  and in case the metric $g_W$ is such that $W$ is an almost round sphere of a radius properly matched to the cosmological constant, it is believed that
the path integral that computes $\Psi_\HH(g_W)$ is dominated by the case that $H$ is a hemisphere,
of boundary $W$, also with an almost round metric.   This contribution is exponentially large as $G\to 0$ (because the classical action of the hemisphere is of order $1/G$ and negative),
and it is believed that contributions from other manifolds with boundary $W$ are exponentially smaller (since the classical action of the hemisphere is more negative than that of
any other critical point of the path integral that computes $\Psi_\HH$).   

This description of $\Psi_\HH$ makes clear that $\Psi_\HH$ is a sort of gravitational version of $\Psi_\dS$.   The maximum entropy state $\Psi_\max$ of de Sitter space is a simple
extension of $\Psi_\dS$ to include the observer, so  we can hope to interpret  an extension of $\Psi_\HH$ to include an observer as a generalization of $\Psi_\max$.   How to include an observer in the
no boundary path integral was already briefly discussed in \cite{CLPW}.   For a clue, we can consider the no boundary path integral that computes $Z=\la\Psi_\HH|\Psi_\HH\ra$.
(We will later divide $\Psi_\HH$ by $\sqrt Z$ to get a normalized version of the no boundary state.)   It is believed that $Z$ should be computed by a path integral over $D$-manifolds
without boundary.  Assuming that the cosmological constant is positive, a round $D$-sphere is a critical point in this path integral, and it is believed that for small $G$ this is the
dominant contribution.   The classical action of a round $D$-sphere is $-A/4G$, where $A$ is the area of the cosmological horizon of de Sitter space, so in a classical
approximation the contribution of this  critical point is $e^{A/4G}$, times a subleading factor that comes from quantum fluctuations around the critical point.   The logarithm
of this path integral was interpreted \cite{GH} as the de Sitter entropy, which is therefore $S_\dS=\frac{A}{4G}+\cdots$, where the subleading corrections (which are of order $\log G$)
comes from the fluctuations around the critical point.  

How can we include an observer in this discussion?      In our model, the observer is described by the action (\ref{obsaction}),
and propagates on a geodesic.  In Euclidean signature, we will denote this geodesic as $\gamma_E$.   If spacetime is a sphere, then $\gamma_E$ will be a great circle on this
sphere.  The circumference of this great circle is $\beta_\dS$.   The action for a observer of energy $m+q$ to propagate for a Euclidean distance $\beta_\dS$
is $\beta_\dS(m+q)$, and this contributes to the integrand of the path integral a factor $e^{-\beta_\dS(m+q)}$.  If we simply integrate this over $q$, we get a factor
$e^{-\beta_\dS m}\frac{1}{\beta_\dS}$.  A localized observer in any sort of semiclassical de Sitter space has $\beta_\dS m\gg 1$, so the factor $e^{-\beta_\dS m}$ is important,
but the factor $1/\beta_\dS$ is a subleading correction that can be included with the other factors that come from quantum fluctuations.    Ignoring such factors, we can
approximate the path integral including the observer as 
\be\label{iggo} Z=\exp\left(\frac{A}{4G}-\beta_\dS m\right). \ee
   Taking the logarithm, we find that the entropy of de Sitter space with an observer of mass $m$ is (according to the logic of \cite{GH})
\be\label{liggo} S_{\dS,\obs}= \frac{A}{4G}-\beta_\dS m. \ee
This is actually a standard result.  Including in the static path an object of mass $m$ (with $m$ small enough that we can ignore back reaction due to the gravity of this object,
as assumed in the preceding discussion) reduces the entropy of the static patch by $\beta_\dS m$.

 \begin{figure}
 \begin{center}
   \includegraphics[width=3.2in]{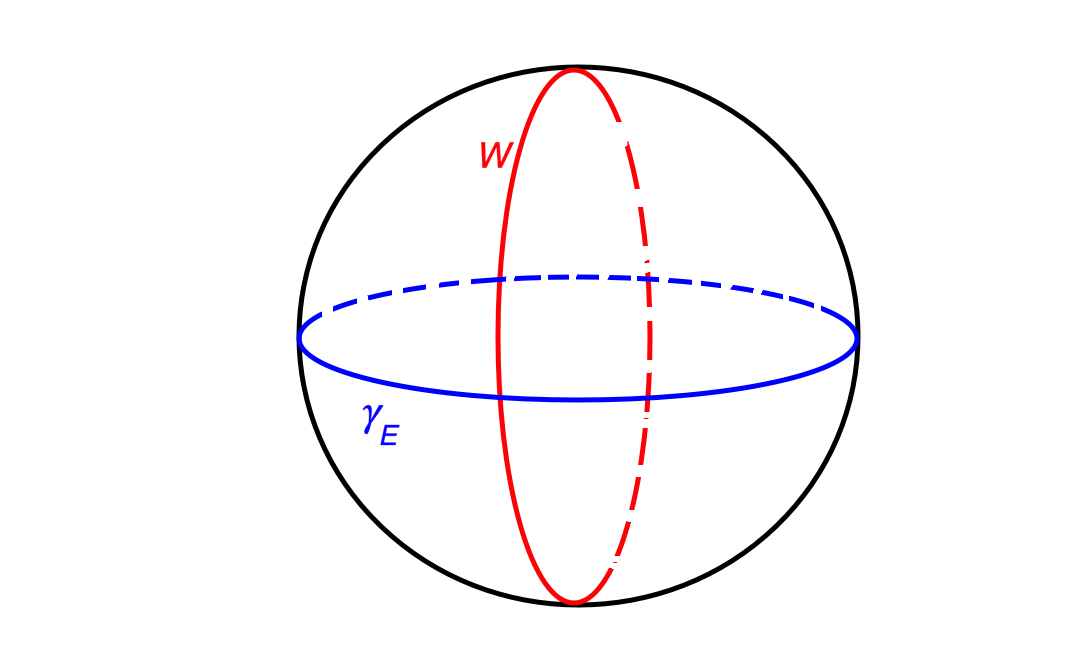}
 \end{center}
\caption{\small \footnotesize{A two-sphere $S^D$ containing an ``equator'' $W\cong S^{D-1}$ orthogonal to a great circle $\gamma_E$.   Drawn is the case $D=2$,
so $W$ is another great circle.   $W$ and $\gamma_E$ intersect at two points and accordingly 
the continuation of $\gamma_E$ to Lorentz signature has two components.}\label{Circles}}
\end{figure} 

To interpret the no boundary state $\Psi_\HH$ in Lorentz signature, the standard procedure is to 
 ``cut'' the Euclidean spacetime on a plane of symmetry $W$ (that is, on the codimension one fixed
point set of a $\Z_2$ symmetry) and then continue to Lorentz signature with $W$ viewed as an initial value surface.   In the absence of the observer, because of the
assumed $\Z_2$ symmetry, 
this gives a real solution in Lorentz signature if the original Euclidean solution is real.   In the case that the Euclidean spacetime is a sphere $S^D$, an appropriate $W$ is
an ``equator'' $W\cong S^{D-1}$.    In the presence of an observer, we want a further condition that $\gamma_E$ continues in Lorentz signature to a real geodesic, which will
be the observer worldline.  To make this true, $W$ must be orthogonal to $\gamma_E$ (fig. \ref{Circles}).  $W$ and $\gamma_E$ intersect at two points, and the continuation
of $\gamma_E$ to Lorentz signature is actually the disjoint union of two timelike geodesics $\gamma$ and $\gamma'$ that are spacelike separated.    (This
is analogous to what happens for an accelerated observer in Minkowski space \cite{Unruh}; the Euclidean orbit is a circle, and its continuation to Lorentz signature is a hyperbola
with two components.)  In fig. \ref{AAA}, if
$\gamma$ is the left edge of the Penrose diagram, then $\gamma'$ is the right edge.  We can think of $\gamma$ as the worldline of the observer that we have been studying
in this article, and $\gamma'$ as the worldline of a second observer who is entangled with the first.   

If we had not integrated over $q$,   we would have written the partition function as $\int_0^\infty \d q e^{A/4G-\beta_\dS m-\beta_\dS q}$ (times additional factors from
quantum fluctuations).
When we ``cut'' on $W$ to divide the sphere into two hemispheres, we associate to each hemisphere the square root of the integrand in this integral or
$e^{A/8G-\beta_\dS m/2 -\beta_\dS q/2}$.   In particular, this gives the no boundary state as a function of $q$:
it is proportional to  $e^{-\beta_\dS q/2}$.   This coincides with the $q$-dependence  of the maximum entropy state
$\Psi_\max$, so we learn that in the context of de Sitter space, the no boundary state coincides with the maximum entropy state, at least to the extent
that they are both defined and understood.   We cannot be sure that either or both of them make sense beyond perturbation theory or if so, that they agree beyond
perturbation theory.   To compute the no boundary partition function $Z=\la\Psi_\HH|\Psi_\HH\ra$ 
 from the no boundary state, we multiply  two factors of $e^{A/8G-\beta_\dS m/2 -\beta_\dS q/2}$,
one from the northern hemisphere and one from the southern hemisphere, or one from the bra and one from the ket, and integrate over $q$ to evaluate the inner product
of the bra and the ket.

 Some puzzles about this setup were described and not entirely resolved in \cite{CLPW}. Those issues will not be repeated here.  

\subsection{A Universal No Boundary State?}\label{universal}

We wish to consider a hypothesis with two parts.

The first part of the hypothesis asserts, roughly, that the no boundary state $\Psi_\HH$, enriched to include the observer, makes sense universally as a state of the observer
algebra $\A_\obs$.   This means that, regardless of the spacetime $M$  in which the observer lives, one can define the expectation value $\la\Psi_\HH|\a|\Psi_\HH\ra$
of an operator $\a\in \h\A_\obs$.

Actually, in section \ref{algebras}, we will slightly refine this hypothesis to say that in general $\Psi_\HH$ is a weight, rather than a state, of $\h\A_\obs$.   Roughly, this means that
$\la\Psi_\HH|\a|\Psi_\HH\ra$ is defined only for a sufficiently nice class of operators in $\h\A_\obs$, somewhat analogous to trace class operators acting on an infinite-dimensional
Hilbert space.  The difference between a state and a weight will not be important in this section.

The second part of the hypothesis is that $\Psi_\HH$ can be regarded as a universal maximum entropy state.

Under these assumptions, we can give a general definition of entropy for
a state of the observer in any spacetime.   Suppose that $\Psi$ is a state of the algebra $\h\A_\obs$ in some spacetime $M$.
Then by our hypothesis, we have two states of $\h\A_\obs$, namely the given state $\Psi$ and the no boundary state $\Psi_\HH$.
In general, the relative entropy between two states of a von Neumann algebra $\h\A_\obs$ is always defined.   If $\h\A_\obs$ is of Type I or Type II, then density matrices
and traces make sense for $\h\A_\obs$, and we can use the familiar definition (\ref{turnof}) of relative entropy.     Assuming our hypothesis about $\Psi_\HH$, it is reasonable
to suspect that $\h\A_\obs$ is always of Type I or Type II regardless of $M$; 
this point will be discussed in section \ref{algebras}.     But relative entropy between two states of a von Neumann algebra
can always be  defined, even if the algebra is of Type III.  For a Type III algebra,\footnote{For a Type II$_\infty$ algebra, traces and density matrices exist, but
there is no natural normalization of the trace.   If one rescales the trace by $\Tr\to e^c \Tr$, one has to rescale all density matrices by $\rho\to e^{-c}\rho$ in order
to preserve the condition $\Tr\,\rho=1$.  This shifts entropies by $S(\rho)\to S(\rho)+c$;  that is one aspect of the fact that the definition of entropy of a state of a Type II
algebra involves an additive renormalization.  But it does not affect the definition (\ref{turnof}) of the relative entropy.  Of course, that is related to the fact that the
relative entropy between two states of a von Neumann algebra can be defined without existence of a trace at all.}
  one has to use a more abstract definition of relative entropy in terms of a certain
relative modular operator \cite{ArakiTwo,WittenReview}.

Under our hypotheses,
we can give a general definition of the entropy of any state of the observer:
\be\label{toggo} S(\Psi)=-S(\Psi|\Psi_\HH). \ee
On the right hand side, $S(\Psi|\Psi_\HH)$ is the relative entropy between the two states $\Psi$ and $\Psi_\HH$ of $\h\A_\obs$.  If $\h\A_\obs$ is of Type I or Type II,
then $\Psi$ and $\Psi_\HH$ can be described by density matrices $\rho$ and $\rho_\HH$, and 
we can restate eqn. (\ref{toggo}) in the form
\be\label{boggo}S(\rho)=-\Tr \rho( \log \rho-\log \rho_\HH).\ee   Otherwise, one has to use a more abstract definition of relative entropy.
The proposal (\ref{toggo}) is natural if it is true that $\Psi_\HH$ can be viewed as a state of maximum entropy and can be regarded as a state of the observer algebra
in any spacetime.   The intuition for $\Psi_\HH$ to be a state of maximum entropy is that it is a sort of global version of $\Psi_\max$, which is a state of maximum
entropy in a particular de Sitter vacuum.   The reason to hope that $\Psi_\HH$ makes sense in any (closed) spacetime is that naively, the recipe to compute it by integrating over
all bulk manifolds with given boundary data seems to be universal.

In one interesting situation, we can show that the definition (\ref{toggo}) gives a sensible answer.   This is the case that the spacetimes that we consider are different de Sitter
spacetimes $M^\alpha$, in a theory that has many different inequivalent de Sitter vacua.  Each $M^\alpha$ has its own Hilbert space $\H^\alpha$, inverse temperature
$\beta_\alpha$ and horizon area $A^\alpha$. 
  Each $M^\alpha$ also has its own maximal entropy state $\Psi_{\max,\alpha}$, with density matrix 
$\rho_{\max,\alpha}={\bf 1}_\alpha$ (here ${\bf 1}_\alpha$ is the identity operator on $\H^\alpha$).    We will assume that the observer Hamiltonian is the same $H_\obs=m+q$
independent of $\alpha$, but this could easily be generalized by letting the coefficients in the action (\ref{obsaction}) depend on a scalar field that has different expectation
values in different de Sitter vacua.   We could also generalize the discussion to allow the possibility that $G$ has different effective values in different vacua.

In the approximation of considering only the spacetimes $M^\alpha$, the no boundary partition function, summed over connected manifolds, can naively be read off
from eqn. (\ref{iggo}):
\be\label{loggo} Z=\sum_\alpha \exp\left(\frac{A^\alpha}{4G}-\beta_\alpha m\right).\ee
This formula should not be taken very literally because it includes exponentially small corrections from manifolds with non-maximal values of $A^\alpha$, but ignores
perturbative corrections to the contribution with maximal $A^\alpha$. 
At any rate, the precise value of $Z$ will not be very important in what follows.

In the no boundary state, the probability that the observer is living in $M^\alpha$ is
\be\label{loggox}p_\alpha=\frac{Z^\alpha}{Z}=\frac{1}{Z}\exp\left(\frac{A^\alpha}{4G}-\beta_\alpha m\right),\ee
where $Z^\alpha$ is the partition function if the observer is in $M^\alpha$, and we used the result eqn. (\ref{iggo}) for $Z^\alpha$. 
If the observer does live in $M^\alpha$, then the no boundary state with the observer present reduces to the
state $\Psi_{\max,\alpha}$, which is the maximum entropy state in that spacetime, as
we saw in section \ref{bothstates}.    This tells us what must be the density matrix of the no boundary state:
\be\label{tolox}\rho_\HH=\frac{1}{Z} \sum_\alpha \exp\left(\frac{A^\alpha}{4G}-\beta_\alpha m\right)\cdot {\bf 1}_\alpha.  \ee
In other words, if the universe is in the state $\Psi_\HH$, then the observer is in $M^\alpha$ with probability $p_\alpha$, and if so the observer experiences a maximum entropy state in that spacetime.

In this particular case, density matrices are available since $\h\A_\obs$ is of Type II in each universe.   Using eqn. (\ref{tolox}) for $\rho_\HH$ along with $\rho_{\max,\alpha}={\bf 1}|_\alpha$,
we can evaluate eqn. (\ref{boggo}):
\be\label{zonko}S(\rho_{\max,\alpha})=\frac{A^\alpha}{4G}-\beta_\alpha m -\log Z. \ee
This is a satisfactory answer.   Up to the overall constant $-\log Z$, it agrees with the expected value of the entropy of the maximum entropy state of $M^\alpha$,
including the area term $A^\alpha/4G$ and the reduction by $\beta_\alpha m$ because of the presence of the observer.  

If instead of putting $M^\alpha$ in the state $\Psi_\max$, we consider a state that is an $\O(1)$ perturbation of $\Psi_\max$ and such that
 the generalized entropy $S_{\rm gen}(\rho)=A/4G+S_{\rm out}$ can be defined,
 the analysis of \cite{CLPW} can be applied and extends eqn. (\ref{zonko}) to get $S=S_{\rm gen}-\log Z$.   Thus at least for this class of spacetimes and states,
entropy defined using our hypothesis about the no boundary
state agrees with the usual generalized entropy, up to a universal additive constant $-\log Z$.  In the derivation in \cite{CLPW},
the $A/G$ term contributed to entropy differences between states, but, since the states considered were $\O(1)$ perturbations of $\Psi_\max$,
they had values of $A/G$ that differ only by $\O(1)$.   In eqn. (\ref{zonko}), the $A/G$ terms makes a contribution  to entropy differences of order $1/G$.

\section{More On The No Boundary State}\label{algebras}

Let $M$ be a spacetime in which the observer may be living.   To be precise, we define $M$ by a solution of the appropriate gravity theory, and a geodesic $\gamma\subset M$
that will be the worldline of the observer.   Then we quantize small fluctuations around the chosen solution to construct a Hilbert space $\H$.  This definition makes sense at least
to all orders of perturbation theory.   The observer algebra $\A_\obs$ can be completed to an algebra $\h \A_\obs$ of operators on $\H$.

In section \ref{unn}, we ask two questions about this setup:

(1) Is there a state in $\H$ that has maximum entropy for $\h\A_\obs$?

(2)  To what extent does the no boundary state $\Psi_\HH$ make sense as a state in $\H$?

Though we will not be able to get firm answers to  these questions, we will motivate the following answers. 
On the first question,
generically there is no state in $\H$ of maximum entropy.
 On the second question, generically $\Psi_\HH$ does not make sense
as an ordinary normalizable  state in $\H$, but it may be that generically $\Psi_\HH$ makes sense as, roughly, an unnormalizable state (more precisely, as a weight
for $\h\A_\obs$).   

In section \ref{tying}, we discuss how these two questions are potentially related to each other and to the ``type'' of the von Neumann algebra $\h\A_\obs$. 

\subsection{Unnormalizable States and Unbounded Entropies}\label{unn}

We will consider two examples of relatively simple spacetimes that are still more complicated than empty de Sitter space.  This discussion will be heuristic and speculative on the
most interesting points.  Then we will be even more speculative about a general picture.   

For our first example, we imagine turning on a scalar or electromagnetic field in de Sitter space.   For definiteness, consider a scalar field $\phi$.  
 Pick a particular
$G$-independent profile $\varphi$ for the scalar field, and consider a one-parameter slice in the space of scalar fields, say $\phi=u \varphi$, with $u$ a real parameter.
Now we want to set $u=u_0$ where $u_0>0$ is large, say of order $G^{-1/2}$, so that turning on $\phi$ with coefficient $u_0$ is an $\O(1/G)$ perturbation, rather than an
$\O(1)$ perturbation, of the original de Sitter space.   However, we can assume that the coefficient of $G^{-1/2}$ is small enough that back reaction on the metric
is not important.   What we have described is then to good approximation a strong scalar field in a background de Sitter space.

Now expanding around the background with $u=u_0$, we can quantize all the small fluctuations and construct a Hilbert space $\H$.     A state  $\chi\in \H$ is a function of 
infinitely many modes, including one that describes a fluctuation in $u$, say with $u=u_0+x$.     Like all the fluctuating modes on which $\chi$ depends, $x$ is supposed
to be of order 1, not order $1/G^{1/2}$.   

Introducing now an observer so that we can hope to define entropy, we can  ask, ``Is there a state in $\H$ with maximum entropy for the observer algebra $\A_\obs$?''
The answer to this question is going to be ``no'' for the following reason.   Since empty de Sitter space has maximum entropy, turning on $\phi\sim u$ has reduced
the entropy of de Sitter space.   We can increase the entropy by making $|u|$ smaller; since $u=u_0+x$ and $u_0>0$, we should take $x<0$. With $u_0\sim G^{-1/2}$
and $x\sim 1$, within any semiclassical picture we always have $|x|\ll |u_0|$ and we can always make the entropy of a state bigger by making $x$ more negative.
So there is no maximum entropy state in $\H$.  

The story for the no boundary state is somewhat similar.   The no boundary state of a free scalar field coupled to a background gravitational field is a Gaussian, since the path integral
over $\phi_H$  in eqn. (\ref{zilno}) is Gaussian in the case of a free scalar field in a background gravitational field.   So we can assume $\Psi_\HH(u)=C\exp(-E u^2)$ with constants $C, E$.   
Now we expand $u=u_0+x$, so viewed as a function of $x$,  $\Psi_\HH(x) =C\exp(-E(u_0+x)^2)$.   This state has two key properties.   With $u_0\sim G^{-1/2}$ and $x\sim 1$, it is extremely small,
exponentially small as $G\to 0$.    But  $\Psi_\HH$ cannot be viewed as a normalizable state in $\H$,   since as long as $u_0\sim G^{-1/2}$ and $x\sim 1$, $\Psi_\HH$ 
grows indefinitely as $x$ becomes
more negative.

It may seem contradictory to say that a state is unnormalizable and also that it is exponentially small.  It means that for given $x$, the state is exponentially small for $G\to 0$, but the dependence on $x$ is such that the state is not normalizable.

Though $\Psi_\HH$ cannot be viewed as a normalizable vector in $\H$,   we were able to write it 
as a function of $x$, so  one might be tempted to think that $\Psi_\HH$ is an unnormalizable state in $\H$.
Of course, by definition, a Hilbert space does not contain unnormalizable states, so the phrase ``unnormalizable state in $\H$'' is problematical.   However,
in von Neumann algebra theory there is a notion of a ``normal weight'' which roughly corresponds to the intuition of an unnormalizable state.  For $\Psi_\HH$ to be a normal
weight\footnote{For a slightly more detailed explanation, see section 4 of \cite{PW}.}  of the algebra $\A_\obs$ acting on $\H$ means in part that there are some positive operators  $\a\in\h\A_\obs$ such that $\la\Psi_\HH|\a|\Psi_\HH\ra<\infty$.
An example of a positive  operator with a finite expectation value in $\Psi_\HH$ is the projection operator on $x\geq -r$, which we will call $\Xi_r$.   Since $\h\A_\obs$ contains any bounded
function of any mode of the $\phi$ field that the observer can measure, and $x$ is the only mode of $\phi$ that is effectively unbounded in $\Psi_\HH$, it is plausible that $\h\A_\obs$
contains a projection operator similar to $\Xi_r$ with a finite expectation value in $\Psi_\HH$.   For $\Psi_\HH$ to be a normal weight, one
 also wants to know that the function $\a\to \la\Psi_\HH|\a|\Psi_\HH\ra$
on positive elements of $\h\A_\obs$ is a limit of increasing functions $\la\Psi_n|\a|\Psi_n\ra$,  where $\Psi_n$, for $n=1,2,3,\cdots$, are normalizable states in $\H$.   Here we can possibly
take $\Psi_n=\Xi_n\Psi_\HH$.

A somewhat similar example is the Schwarzschild de Sitter (SdS)  solution, describing a black hole in de Sitter space.   This solution depends on a free parameter, the black hole
horizon area $A_\BH$.  There is also a canonical conjugate of $A_\BH$, which is a sort of global time-shift mode.   The entropy is a decreasing function of $A_\BH$; it is
minimized when $A_\BH$ has the largest possible value (this corresponds to the Nariai solution, with $A_\BH$ equal to the area of the cosmological horizon) and maximized
in the rather singular limit $A_\BH\to 0$, where the topology changes as the two sides of the black hole become disconnected.    Now consider an SdS solution with a typical value $A_\BH=A_0\sim \beta_\dS^{D-2}$.   We can define a Hilbert space $\H$ that
describes $\O(1)$ fluctuations around this  SdS solution including fluctuations in $\A_\BH$, say with $A_\BH=A_0+y$, where $y\ll A_0$.  

The parameter $y$ will play a role similar to that played by $x$ in the previous
example.   There is no maximum entropy state in $\H$, because in the context of the Hilbert space $\H$, we can always increase the entropy by making $y$ more negative.
What about $\Psi_\HH$?
Heuristically, because the entropy of the SdS solution is less than that of empty de Sitter space, one would expect that the no boundary state $\Psi_\HH$ is exponentially
small in $\H$.   But because that entropy increases as $y$ becomes more negative, one would expect that $\Psi_\HH$ is unnormalizable as a state in $\H$.
Similarly to the discussion of the previous example, it is plausible that $\Psi_\HH$ can be interpreted as a normal weight of the algebra $\h\A_\obs$ acting on $\H$.
A projector on $y\geq-r$ could play the role of $\Xi_r$ in the previous case.

The relevant difference between the two examples is the following.   In the first example, the deformation to a maximum entropy state by turning off the scalar field perturbation
is a completely smooth and straightforward process classically.   The second example is less straightforward classically, since the  limit
$A_\BH\to 0$ is not really a smooth classical limit.   Presumably  when $A_\BH$ gets sufficiently small, the semiclassical picture breaks down and the black hole evaporates,
disconnecting the two sides of the black hole and replacing a Cauchy hypersurface $S^{D-2}\times S^1$ with $S^{D-1}$.
This is a relatively exotic form of spacetime topology change, though one about which we have some inkling.

What do we think happens in a generic closed universe $M$?  Consider a Hilbert space $\H$ that describes small fluctuations around some classical solution on $M$.
 We have very little idea of the behavior of $\Psi_\HH$ as a vector in $\H$, because except in a few cases, a stable Euclidean
solution that is a candidate to dominate the evaluation of $\Psi_\HH$ is not known.   We expect that $\Psi_\HH$ is exponentially small in $\H$ because  presumably the entropy of $M$
 is smaller than that of empty de Sitter space.   Is $\Psi_\HH$  normalizable as a vector in $\H$? Possibly it is, but there is no obvious reason to think so.  Plausibly the two
examples that we discussed are typical and that $\Psi_\HH$ grows exponentially in some directions in field space.  It seems much more likely for $\Psi_\HH$ to be a normal weight 
of $\h\A_\obs$ than a normalizable state.   As in the SdS case, the directions in field space in which $\Psi_\HH$
grows exponentially may bring us towards topology-changing transitions to a higher entropy state,  though  these may  generically be highly nonclassical transitions of which we have no idea.
In that case, one would expect that there is no maximum entropy state in $\H$.  Maximizing the entropy would require moving in the direction of some topology-changing transitions.
   
\subsection{The ``Type'' Of The Von Neumann Algebra}\label{tying}

Though the observer algebra $\A_\obs$ is not an algebra of Hilbert space operators, once we pick a spacetime $M$ that the observer lives in, we can
refine and complete $\A_\obs$ to a von Neumann algebra $\h\A_\obs$.   The question we will ask in this section is what is the ``type'' of this von Neumann algebra.
For our purposes, the relevant types of von Neumann algebra are as follows (for more detail, see \cite{Sorce,Witten}).  We consider only algebras that are ``factors,'' in the sense
that their center consists only of $c$-numbers.

A Type I algebra has an irreducible representation in a Hilbert space $\H$.   This is the usual situation in ordinary quantum mechanics and the structures here are familiar.

A Type II algebra has no irreducible representation in a Hilbert space, so for such an algebra there is no notion of a quantum microstate.   However, a Type II algebra
does have a trace, and therefore density matrices and entropies can be defined for a state of a Type II algebra.   Physically, the entropy of a state of a Type II algebra
is a renormalized entropy, from which an infinite constant (independent of the state) has been subtracted.

A Type III algebra has no irreducible representation in a Hilbert space, and it also has no trace and no notion of a density matrix or entropy.

For an observer in a universe $M$, closed or open, it is definitely possible to have a Cauchy hypersurface $W$ no part of which is hidden by either a past or future horizon.   In such a case, one expects that the algebra $\h\A_\obs$ will be a Type I algebra, the algebra of all operators on $\H$.   However, if there is no Cauchy hypersurface
in the region causally accessible to the observer, one may expect that the observer does not have access to quantum microstates and that $\h\A_\obs$ will be of Type II or Type III.

For the static patch in de Sitter space, one can convincingly argue that $\h\A_\obs$ is of Type II.   In a generic spacetime, this is rather unclear.

A key difference between a Type II algebra and one of Type III is that for a state of a Type II algebra, but not for a state of a Type III algebra, there is a reasonable notion of
entropy.   With our hypothesis that the no boundary state $\Psi_\HH$ is a universal state of maximal entropy, we have a general definition of entropy in terms of relative
entropy between a given state and the no boundary state.   Therefore, if this hypothesis is correct, one may suspect that $\h\A_\obs$ is always of Type II.

An obstruction to this idea has been simply that a Type II algebra has a trace, and for an observer in a generic spacetime, it has been quite difficult to imagine how a trace
could possibly be defined.   However, the hypothesis concerning the universal nature of the no boundary state gives a possible answer.

For our purposes, there are two types of Type II algebra.\footnote{These are the two types of ``hyperfinite'' Type II algebra.  A hyperfinite algebra is one that can be approximated
by finite dimensional matrix algebras.   If one relaxes the assumption of hyperfiniteness, the classification of Type II algebras is much more involved.}   A Type II$_1$ algebra $\A$
has a representation in a Hilbert space $\H$ such that there is a ``tracial'' vector $\Psi_{\rm tr}\in\H$.  A tracial vector is a vector with the property that the trace in the algebra
is the expectation value in that state:
\be\label{weggo}\Tr\,\a=\la\Psi_{\rm tr}|\a|\Psi_{\rm tr}\ra,~~~\a\in\A.   \ee  One usually normalizes the tracial vector by $\la\Psi_{\rm tr}|\Psi_{\rm tr}\ra=1$, ensuring that
$\Tr \,1=1$.
As explained in section \ref{maxent},  a tracial vector automatically defines a state of $\A$ of maximum entropy.

A Type II$_\infty$ algebra, roughly speaking, has the same property except that $\Psi_{\rm tr}$ is 
unnormalizable.   To be more precise, $\Psi_{\rm tr}$ is a normal weight of the algebra, a concept briefly introduced in section \ref{unn}.   
In a Type II$_\infty$ algebra, because $\Psi_{\rm tr}$ is unnormalizable,  we cannot 
normalize it by a condition like $\la\Psi_{\rm tr}|\Psi_{\rm tr}\ra=1$, and in fact there is no natural way to normalize the trace in a Type II$_\infty$ algebra. 
(There is an obstruction: the algebra has a group of outer automorphisms that rescales the trace.) Moreover, in a Type II$_\infty$
algebra, the trace is not defined for all elements of the algebra, since for example if $\Psi_{\rm tr}$ is unnormalizable, then eqn. (\ref{weggo}) implies that $\Tr\,1=\infty$.

How can one possibly define a trace in a generic spacetime?  If one is willing to hypothesize that the no boundary state $\Psi_\HH$ can be defined for any closed universe,  
then this suggests that $\Psi_\HH$ is itself the tracial state:
\be\label{leggo}\Tr\,\a =\la\Psi_\HH|\a|\Psi_\HH\ra.\ee
If so, then $\h\A_\obs$ is of Type II$_1$ in the (possibly very exceptional) case that $\Psi_\HH$ is normalizable in a given spacetime, and Type II$_\infty$ otherwise.
There is a maximum entropy state in a given spacetime if and only if $\Psi_\HH$ is normalizable in that spacetime.   
This is in reasonable agreement with the heuristic discussion in section \ref{unn}.

Eqn. (\ref{leggo}) makes more sense if it is true that $\Psi_\HH$ is unnormalizable in a given spacetime, because in that case, the trace is defined only for operators
that in some way cancel or project out the divergence in $\Psi_\HH$ in the given spacetime.   Not understanding $\Psi_\HH$ in a generic spacetime, we do not 
understand what are the operators
for which we should define a trace, and that helps explain why it is hard to see that a trace exists.  If $\Psi_\HH$ is normalizable for a given spacetime, one expects to 
define a trace valid for all operators in $\h\A_\obs$, and one could hope that such a trace would be more visible.

Probably the best that we can say about eqn. (\ref{leggo}), apart from the fact that it can be verified for the static patch in de Sitter space and possibly in a few other special
cases,  is that it is difficult to disprove this conjecture, because we know so little about the no boundary state in a generic spacetime.

Going back to the case of a closed universe in which $\A_\obs$ is of Type I, what then plays the role of the no boundary state?    Let $M$ be such a spacetime with
Hilbert space $\H_M$.   To get a sensible answer, we have to interpret the restriction of $\rho_\HH$ to $\H_M$ -- that is, to the case that the observer is in $M$ --
as $\rho_\HH|_{\H_M}=\frac{1}{Z}{\bf 1}_{\H_M}$.    This formula makes sense in the spirit of the proposal we are exploring because it is formally small -- as $Z$ is exponentially
large -- but its trace is divergent, so it is not the density matrix of a normalizable state. (Rather, the function $\a\to \Tr\,\a \rho_\HH$ is a weight of the Type I algebra of
all bounded operators on $\H_M$.)   With this proposal for $\rho_\HH|_{\H_M}$ and
with $\rho$ being any density matrix on $\H_M$,  the general formula (\ref{toggo}) for entropy gives
\be\label{ilfox} S(\rho)=-\Tr\,\rho\log \rho -\log Z,\ee
which is the standard answer up to the universal additive constant $-\log Z$.  That constant appears because we have defined entropy relative to the maximum entropy state
$\Psi_\HH$.   Perhaps the claim that $\rho_\HH|_{\H_M}=\frac{1}{Z}{\bf 1}_{\H_M}$ whenever the observer has causal access to a complete Cauchy hypersurface in a closed
universe can shed light on
a general understanding of the no boundary state.   Note that this formula for the density matrix of $\Psi_\HH$ implies that as a Hilbert space vector, $\Psi_\HH$ can be naturally taken
 to live
in the Hilbert space of the disjoint union of $M$ with a time-reversed conjugate of itself.

\section{Spacetimes That Do Have Asymptotic Observers}\label{AAdS}

Up to this point, we have considered observers who actually live in the spacetime under study.   As explained in the introduction, one
 motivation for this choice is 
that we ourselves are in that situation; another motivation is that in a closed universe and in many standard cosmological models, there is no reasonable notion of an 
observer who can look at spacetime from  outside.   

However, it is also interesting to consider an asymptotically flat or asymptotically Anti de Sitter (AAdS) spacetime, in which there
can be an asymptotic observer at infinity, essentially looking at spacetime from outside.    In such cases, one can consider asymptotic observables without explicitly introducing an observer
who is making them, and this is the standard practice.  

In particular, 
as in \cite{LL,LL2} and various later papers \cite{GCP,CPW}, in the context of AdS/CFT duality, it is interesting to define an algebra generated by  single-trace operators of the boundary
theory  in the large $N$ limit.   Here we consider operators defined on a particular asymptotic boundary -- where an asymptotic observer may be living -- in a spacetime that
may or may not have additional asymptotic boundaries.   For convenience, we will assume that the boundary theory is a four-dimensional gauge theory; the statements
have straightforward modifications for other cases.

What has been studied in the recent literature is an algebra of single-trace operators normalized so that their connected two-point functions -- or equivalently,
their commutators -- are of order 1. 
 Assuming the action is normalized as $I=N \Tr\, L$, where $L$ is a gauge-invariant polynomial in the fields and their derivatives (with no explicit
dependence on $N$), the single-trace operators with two-point functions and commutators of order 1 are generically of the form $\O=\Tr\,W$, where again $W$ is a gauge-invariant function
with no explicit $N$-dependence.    However, operators of this form do not have large $N$ limits.   For example, at inverse temperature $\beta$, their thermal expectation values
$\la \O\ra_\beta$ are of order $N$.  One way to define operators that have a large $N$ limit is to subtract the expectation values of the single-trace operators.
For example, one can consider the operators $\O-\la\O\ra_\beta$, which have a large $N$ limit at inverse temperature $\beta$.   These operators generate
an algebra that has been studied fruitfully, but it is not background-independent.  Above the Hawking-Page transition, it describes $\O(1)$ perturbations of a black
hole at inverse temperature $\beta$.   Background independence was lost by subtracting the thermal expectation values
at a particular temperature.  

As an alternative, one can define operators that have a large $N$ limit by dividing the single-trace operators by an extra factor of $N$.
Thus, one considers operators of the general form $\W=\frac{1}{N}\Tr\,W$.   These operators have large $N$ limits,
and likewise any function of these operators $\F(\W_1,\W_2,\cdots)$ (with no explicit dependence on $N$) has a large $N$ limit.   
The algebra $\A$ generated by such functions is background-independent, since in defining it we have made no
 choice of background.  This algebra 
 makes sense in the
large $N$ limit and  in the $1/N$ expansion. But at $N=\infty$, this algebra is commutative, since dividing by an extra
factor of $N$ 
gives operators that  have commutators of order $1/N^2$:
\be\label{commu}[\W_i,\W_j]=\frac{\i}{N^2}\P_{ij}+\O(1/N^4). \ee
Here $\P_{ij}$ is a function of the $\W$'s, with no explicit $N$-dependence. (We have included a factor of $\i$ so that if the $\W_i$ are hermitian, then 
the $\P_{ij}$ are also hermitian.)  In general, the $\P_{ij}$ are highly nonlinear functions of the $\W$'s.   
Of course the commutator $[\W_i,\W_j]$ satisfies the Jacobi identity.   But if we consider only the terms of $\O(1/N^2)$ in the commutators, there is a further identity
\be\label{ommu}[\W_i,\W_j\W_k]=\frac{\i}{N^2} \left( \P_{ij}\W_K+\P_{ik}\W_j\right)+\O(1/N^4), \ee
since a connected three-point function of the operators $\W_i,\W_j,\W_k$ is of order $1/N^4$.
To formalize the idea of keeping only the terms of order $1/N^2$, let us define, for any functions $\F,\G$ of the single-trace operators,
\be\label{pommu}\{\F,\G\}=\lim_{N\to\infty}(-\i N^2) [\F,\G]. \ee
For example, 
\be\label{zomu} \{\W_i,\W_j\}=\P_{ij}. \ee  Obviously these brackets are antisymmetric.
The Jacobi identity for commutators implies a Jacobi identity for these brackets; for  $\F,\G,\K\in \A$, 
\be\label{omu}\{\F,\{\G,\K\}\}+\{G,\{\K,\F\}\}+\{\K,\{\F,\G\}\}=0.\ee
Eqn. (\ref{ommu})  implies that
\be\label{nomu}\{\W_i,\W_j\W_k\}=\{\W_i,\W_j\}\W_k+\{W_i,\W_k\}\W_j .\ee
This generalizes to
\be\label{lomu} \{\F,\G\K\}=\{F,\G\}\K+\{\F,\K\}\G. \ee
A commutative algebra with an antisymmetric bracket that satisfies the Jacobi identity  (\ref{omu}) and the identity (\ref{lomu}) is called a Poisson algebra,
so the large $N$ limit of $\A$ is a Poisson algebra.   Of course, in perturbation theory in $1/N^2$, $\A$ is deformed to be an associative but noncommutative algebra.
To exhibit the dependence of the algebra $\A$ on $N$, we will denote it as  $\A_{1/N^2}$, so $\A_0$ is a commutative Poisson algebra, and $\A_{1/N^2}$ is noncommutative
for $1/N\not=0$.

Why is the large $N$ limit a Poisson algebra?   The bulk dual of the theory under discussion has a classical phase space,
consisting of classical solutions of the relevant gravity or string theory.   In fact, it has   many possible classical phase spaces, differing by the possible existence (and geometry
and topology)
of additional asymptotic boundaries apart from the one where we are defining the algebra, and by the bulk topology that is assumed.   Let  $S$ be the set
of possible bulk phase spaces, and let us denote those
phase spaces as $\M_\lambda$, $\lambda\in S$.  A point on any of the $\M_\lambda$ determines a classical solution of the 
bulk gravity or string theory, and the asymptotic behavior of the bulk fields in
this solution determines the expectation values of the $\W$'s.  To say this differently, in the large $N$ limit, the $\W$'s are functions on $\M_\lambda$ (for each choice of $\lambda$).   
As classical phase spaces, the $\M_\lambda$ have symplectic structures which enable one to define Poisson brackets.   
Note that on any classical phase space, the Poisson brackets of functions $f,g,k$ satisfy
\be\label{lomuu}\{f,gk\}=\{f,g\}k+\{f,k\}g, \ee in perfect parallel with (\ref{lomu}).
 So the functions on any classical phase space form a Poisson algebra.\footnote{\label{Poisson}More generally, a manifold 
 $M$ with a Poisson bracket $\{f,g\}=\alpha^{ij}\partial_i f\partial_j g$
   that satisfies the Jacobi identity, where $\alpha^{ij}$ is an antisymmetric tensor field on $M$,
is called a Poisson manifold (and $\alpha^{ij}$ is called a Poisson tensor). Such a Poisson bracket automatically satisfies eqn.  (\ref{lomuu}), so the functions on such a manifold form a Poisson algebra.   If $\alpha^{ij}$ is invertible, then the Jacobi identity implies that 
its inverse is a symplectic form $\omega_{ij}$, and in that
case $M$ is a symplectic manifold -- a classical phase space.  A simple example of a Poisson manifold with non-invertible Poisson tensor is a Lie algebra $\mathfrak g$ with Poisson brackets
$\{x_a,x_b\}=f_{ab}^c x_c$, where $f_{ab}^c$ are the structure constants of $\mathfrak g$. }
In fact, in the AdS/CFT correspondence, the $1/N^2$ expansion of the boundary theory matches the expansion of the bulk theory in powers of $G\hbar$, so the Poisson brackets of
the bulk theory, which are the leading term in $G\hbar$ of  the commutators of bulk operators, map to the leading term in $1/N^2$ of the commutators of single-trace operators.

In deformation quantization, one is given a classical phase  $\M$ or a more general Poisson manifold as in footnote \ref{Poisson}.  
The goal is to deform the commutative algebra $\A$ of functions on $\M$
to an associative but noncommutative algebra $\A_\hbar$, order by order in a parameter $\hbar$, with $[f,g]=\i\hbar\{f,g\}+\O(\hbar^2)$, and with $[f,g]$, in order $\hbar^k$,
being defined locally in terms of derivatives of $f$ and $g$ up to $k^{th}$ order.   
We are in this setting except that our Poisson algebra is associated with not one Poisson manifold but many.      
The problem of deformation quantization (at least in the usual case of a single Poisson manifold) has a general solution that is unique up to a certain
kind of equivalence \cite{Sternheimer,WL,Fedosov,Kontsevich,CF,CF2}.   In our problem of the AdS/CFT correspondence, we do not need to invoke general theorems
to know that the problem has a solution, since we know that the quantum
theory under study exists for every integer $N$ and that it has an asymptotic expansion of an appropriate form near $N=\infty$.   However, it is worth mentioning that one very 
interesting approach to deformation quantization \cite{Kontsevich,CF,CF2}  involves a path integral on a disc with the algebra elements inserted on the boundary of the disc.   
As in two-dimensional models of a black hole such as JT gravity, the path integral on the disc naturally produces an algebra.
The rotation symmetry of the disc enables one to endow this algebra with a trace if the Poisson manifold is symplectic (that is, if the Poisson tensor is invertible).
However,  this algebra does not have a natural Hilbert space representation, or more precisely, it does not have a natural
  ``one-sided'' Hilbert space representation that could represent quantization of a single copy\footnote{When the algebra
has a trace, there is a natural two-sided Hilbert space, as in the black hole case:  the algebra itself can be regarded as a Hilbert space, with $\la \a,\b\ra =\Tr\,\a^\dagger\b$.
This gives a Hilbert space with the algebra acting on itself by left multiplication, and commuting with a similar algebra acting on the right. This Hilbert space can be
interpreted as representing quantization of the product of two copies of $M$.  Of course, in some cases, such as an example discussed below, a natural one-sided Hilbert space
does exist (for certain values of $j$).   This is not always the case and defining a natural one-sided Hilbert space is beyond the scope of deformation quantization.} of $M$.
However, if one picks a point  $p\in M$, then expanding around $p$, one can construct a Hilbert space $\H_p$ on which the algebra acts. 
We will give an example shortly.

From the standpoint of the $1/N$ expansion, the algebra $\A_{1/N^2}$ that we get by deforming the large $N$ Poisson algebra $\A$ order by order in $1/N^2$ is somewhat analogous
to the observer algebra $\A_\obs$ that we have studied in the bulk of the present paper.   It does not have any distinguished Hilbert space representation that can
be defined in the $1/N$ expansion.  However
any choice of a point in any one of the classical phase spaces $\M_\lambda$ determines a Hilbert space representation of $\A_{1/N^2}$.   Indeed, a point $p\in M_\lambda$
determines a classical solution of the bulk gravity or string theory.   Expanding around this point and quantizing the small fluctuations, we get a Hilbert space that
makes sense order by order in perturbation theory.   In the boundary theory, this corresponds to a Hilbert space that makes sense order by order
in $1/N$ and provides a representation of $\A_{1/N^2}$.      Like $\A_\obs$, $\A_{1/N^2}$ is not a von Neumann algebra in any background
independent sense, but once one picks a Hilbert space representation, one can complete it to get a von Neumann algebra.   

A difference between the two cases is that in AdS/CFT, we expect  that $\A_{1/N^2}$ can be 
defined nonperturbatively, in the sense that it is possible to take $N$ to be a positive integer and thus
to assign a numerical value to $1/N^2$ rather than treating it as a formal variable.   
By contrast, the idea of an eternal observer in spacetime is an idealization. The best we can say about $\A_\obs$ is that it makes sense to all orders of perturbation theory;
the precise limitation on the validity of $\A_\obs$ is not clear.   However, in AdS/CFT, it is quite plausible that semiclassical bulk notions of spacetime and causality are
not sharply defined in the nonperturbative theory in which $1/N^2$ is set to  a numerical value.  These notions may  make sense only asymptotically in $1/N^2$.
  The boundary algebra related to a semiclassical spacetime would then be
$\A_{1/N^2}$ with $1/N^2$ treated as a formal variable, sharpening the analogy with $\A_\obs$.   

We will conclude by describing an elementary example of deformation quantization, in the hope that this will make some things clearer.     
The phase space is a two-sphere $M$ parametrized by real variables $x_1,x_2,x_3$ with
\be\label{phasesp}x_1^2+x_2^2+x_3^2=1.\ee
We take the symplectic structure 
to be
\be\label{symp}\omega =(j+1/2) \frac{\d x_1\d x_2}{x_3}. \ee
We could choose the coefficient here to be $j$ rather than $j+1/2$; we will be expanding in $1/j$, which is essentially equivalent to expanding in $1/(j+1/2)$.
 The formulas will look nicer with the choice we have made.
The symplectic form in eqn. (\ref{symp}) is ${\rm SO}(3)$-invariant, though not manifestly so.  It has been normalized so that
\be\label{intm}\int_M\omega =4\pi( j+1/2),\ee
which is an integer multiple of $ 2\pi$, making quantization  possible, if and only if $j\in\frac{1}{2} \Z$.  We can orient $M$ so that $j$ is nonnegative.
Quantizing a sphere with a symplectic form whose integral is $2\pi k$, we expect to get a Hilbert space of dimension $k$,
thus with angular momentum such that $2j+1=k$.   Thus, we anticipate that if $j$ is a half-integer, quantization of $M$ will give a Hilbert space
in the spin $j$ representation of $\SU(2)$.

This symplectic form can be derived from a Lagrangian that is also proportional to $j+1/2$.  Thus $j$ (or $j+1/2$) plays
the role of $N^2$ or $1/\hbar$.   The Poisson brackets derived from the symplectic form are 
\be\label{pois}\{x_i,x_j\}=\frac{1}{j+1/2}\epsilon_{ijk} x_k. \ee 
So in deformation quantization, we want to promote the $x_i$ to operators $\x_i$ that will obey $[\x_i,\x_j]=\frac{\i}{j+1/2}\epsilon_{ijk}\x_k+\O(1/j^2)$ and also obey a relation
that will coincide with the classical relation (\ref{phasesp}) up to a correction of order $1/j$.   In this simple example, we can just write down the answer.   No 
correction is needed to the commutation relations, but there is a $1/j^2$ correction to the classical relation $\sum_i x_i^2=1$.  The algebra relations are 
\be\label{answer} [\x_i,\x_j]=\frac{\i}{j+1/2}\epsilon_{ijk} \x_k,~~~ \x_1^2+\x_2^2+\x_3^2 =\frac{j(j+1)}{(j+1/2)^2}=1-\frac{1}{(2j+1)^2}.\ee
Alternatively,\footnote{We could also reparametrize $j$ by $j\to j+c_0+c_1/j+\cdots,$ with constants $c_0,c_1,\cdots$, and this  would still give a valid solution of the problem
of deformation quantization.  By making such a reparametrization of $j$ and also rescaling the $\x_i$, one could entirely remove the deformation of the classical algebra.
However, the choice we have made is more natural in what follows.}  we could rescale the $\x_i$ by a factor $(1-\frac{1}{(2j+1)^2})^{-1/2}$, and then there would be 
a $1/j^3$ correction to the commutation relations, but no correction to the classical
relation $\sum_i\x_i^2=1$.      We will denote this algebra defined by the relations (\ref{answer}) as $\A_{1/j}$.

It is  quite familiar that when $j$ is a non-negative half-integer, $j\in \frac{1}{2}\Z_{\geq 0}$, this algebra has a representation in a Hilbert space $\H_{(j)}$ of dimension $2j+1$.  However, it is not possible to define a large
$j$ limit of $\H_{(j)}$ 
in an $\SO(3)$- or $\SU(2)$-invariant way.   So the algebra $\A_{1/j}$ defined in perturbation theory in $1/j$, with $1/j$ regarded as a formal variable, does not have a natural
Hilbert space representation.   But once we choose a point $p\in M$, by expanding around $p$, we can define a Hilbert space $\H_p$ that does have a large $j$
limit and on which $\A_{1/j}$ acts, order by order in $1/j$.   In explaining the construction of $\H_p$,
because of the rotation symmetry of the sphere, it does not matter which point we pick.   We will take $p$ to be the point $(x_1,x_2,x_3)=(0,0,1)$.   This means that in the Hilbert space $\H_p$, $x_3$ will be of order $1$, but $x_1$ and $x_2$
will vanish in the large $j$ limit; in fact, they are of order $1/j^{1/2}$.  The Hilbert space $\H_p$ can be constructed without picking a Hamiltonian, but the construction is
possibly more obvious if one picks a Hamiltonian to organize the states.   A convenient Hamiltonian is $H=-x_3$, chosen so that classically its minimum is the point $p=(0,0,1)$
about which we want to expand.   We can solve the  classical relation (\ref{phasesp}) with $x_3=(1-x_1^2-x_2^2)^{1/2}$, so the Hamiltonian
is
\be\label{hamton}H=-(1-x_1^2-x_2^2)^{1/2}=-1+\frac{1}{2}(x_1^2+x_2^2)+\O((x_1^2+x_2^2)^2) \ee
and the Poisson brackets are
\be\label{pbr}\{x_1,x_2\}= \frac{1}{j+1/2} (1-x_1^2-x_2^2)^{1/2}=\frac{1}{j}+\O((x_1^2+x_2^2)/j,1/j^2). \ee
From these formulas, one sees that in the large $j$ limit, $\sqrt j x_1$ and $\sqrt j x_2$ are canonically conjugate variables, and the Hamiltonian is a harmonic oscillator
Hamiltonian.  One can systematically construct perturbation theory in $1/j$ about this starting point.

That perturbative construction could be carried out on any phase space $M$, expanding around any point $p\in M$,  but in the particular case that $M$ is a two-sphere with $\SO(3)$-invariant symplectic form, 
 we can describe the Hilbert space $\H_p$ exactly, not just in perturbation theory in $1/j$.    In the Hilbert space $\H_{(j)}$, the operator 
 $j\x_3$ is an angular momentum generator with eigenvalues 
$j,j-1,j-2, \cdots, -j$.   $\H_p$ is going to be a large $j$ limit of $\H_{(j)}$, with the limit taken in such a way that all states have eigenvalues of $j \x_3$ close to the maximum.
To accomplish this, we simply declare that $\H_p$ has a basis consisting of the eigenstates of $j\x_3$ with eigenvalue $j-n$, where $n$ is kept fixed while $j\to\infty$.
In this way, we define a Hilbert space in which there is a highest weight vector for the ${\rm U}(1)$ subgroup of $\SU(2)$ generated by $j\x_3$, but no lowest weight vector
for that subgroup and no highest or lowest weight vector for any other ${\rm U}(1)$ subgroup of $\SU(2)$.   For any choice of $p\in M$, we can similarly define a Hilbert space
$\H_p$ that has a highest weight vector precisely for the subgroup of $\SU(2)$ that leaves $p$ fixed.   Each of these
Hilbert spaces furnishes a representation of $\A_{1/j}$, order by order in perturbation theory.

Although the algebra $\A_{1/j}$ does not have a Hilbert space representation that has a large $j$ limit, it does have a trace that has a large $j$ limit.   This trace is completely
determined on polynomials in the algebra generators $\x_i$ by the condition that it is $\SU(2)$-invariant and that $\Tr\,1=1$.   $\SU(2)$ invariance implies that
$\Tr\,\x_i=0$ for all $i$ and that $\Tr\,\x_i\x_j=C\delta_{ij}$ for some constant $C$.   The constant can be determined by using the relations  (\ref{answer}): $C=
\frac{1}{3}\left(1-\frac{1}{(2j+1)^2}\right).$
Similarly $\Tr\,\x_i\x_j\x_j$ must be $C'\epsilon_{ijk}$ for some constant $C'$, which using the relations in the algebra and the fact that
$\Tr\,1=1$ can be found  to be $C'=\frac{1}{6}\left(1-\frac{1}{(2j+1)^2}\right)$.
Continuing in this way, it is not difficult to see by induction that the trace of any polynomial in the $\x_i$ is uniquely determined by the relations in the algebra together with $\SU(2)$
invariance and the condition $\Tr\,1=1$.   Moreover, one can show that the trace of a polynomial in the $\x_i$ of degree at most $2n$ is a polynomial in $1-\frac{1}{(2j+1)^2}$ of degree at most $n$.

For $j\in\frac{1}{2}\Z_{\geq 0}$, the algebra $\A_{1/j}$ has a completely natural representation in the Hilbert space $\H_{(j)}$.  Since it was uniquely determined by the symmetries
and a normalization condition,
 the trace constructed in the last paragraph coincides for these values of $j$
with $1/(2j+1)$ times the ordinary trace in the Hilbert space $\H_{(j)}$.   However, the trace as defined in the last paragraph makes sense for any complex  $j$ except for a pole at $j=-1/2$.  
One can ask for what values of $j$  the trace is positive, meaning that $\Tr\,\a^\dagger\a\geq 0$ for any element $\a\in \A_{1/j}$.   This is certainly true for $j\in \frac{1}{2}\Z_{\geq 0}$, since then the trace
is just a positive multiple of the trace in $\H_{(j)}$.   It is also true that the trace is positive in perturbation theory in $1/j$, where $j$ is understood to be real, 
in the sense that for any given $\a$,  
$\Tr\,\a^\dagger \a>0$ in the large $j$ limit and therefore also in perturbation theory in $1/j$.   In fact, a stronger statement is true:  the large $j$ limit of $\Tr\,\a^\dagger\a$
is the integral over $M$ of the classical function that is the large $j$ limit of $\a^\dagger \a$, divided by $4\pi$.   We will leave it to the interested reader to try to prove that.
However, if we set $j$ to a real value that is not a half-integer, the trace is not positive.\footnote{Apart from half-integer $j$, the
 trace is positive if $j=-1/2+\i s$, with $s$ real.}  To see this, let $\a=(\x_1+\i \x_2)^n$.   Then $\a$ annihilates 
$\H_{(j)}$ if $2j<n$.   So  $\Tr\,\a^\dagger \a$ has $n$ zeroes at $j=0,1/2,\cdots, (n-1)/2$.     Since $\Tr\,\a^\dagger \a$ is a polynomial in $1-\frac{1}{(2j+1)^2}$ of degree at most $n$,
it has at most $2n$ zeroes.  Moreover the set of zeroes is invariant under $j\leftrightarrow -1-j$, so at most $n$ of them are nonegative.  Therefore
the  $n$ zeroes we know about at non-negative values of $j$ are all simple zeroes and are all the zeroes at non-negative $j$.   
Since $\Tr\,\a^\dagger\a>0$ for sufficiently large $j$, it is negative in the
region $(n-2)/2<j<(n-1)/2$, between the two largest zeroes.   Since we can make this argument with any choice of $n$, we learn that if we set $j$ to a non-negative numerical value,
the trace is only positive for $j\in \frac{1}{2}\Z_{\geq 0}$.  

Going back to AdS/CFT, in that context we do not expect the Hilbert space above the Hawking-Page transition to have a large $N$ limit, or any sort of regular behavior
beyond whatever follows from the fact that thermodynamic functions (broadly construed to include certain averaged correlation functions) have a smooth behavior for large $N$.
So $\A_{1/N^2}$ is not expected to have a natural Hilbert space representation that makes sense in the $1/N$ expansion, but it has such a representation for any choice of a point in one
of the phase spaces.  It is interesting to speculate that, similarly to $\A_{1/j}$, $\A_{1/N^2}$ can possibly be analytically continued to complex values of $N$.   If there is something
that plays for $\A_{1/N^2}$ the role that we have conjectured the no boundary state to play for $\A_\obs$, it is likely the infinite temperature limit of the thermofield double state.

\vskip1cm
 \noindent {\it {Acknowledgements}}  
 I thank Dong-su Bak, D. Marolf, G. Penington, and L. Susskind for discussions.    Research supported in part by NSF Grant PHY-2207584.
 \bibliographystyle{unsrt} 

\end{document}